%% file: source_v2.tex
\documentclass[prb,aps,reprint,superscriptaddress,footinbib]{revtex4-1}
\usepackage[latin1]{inputenc} 
\usepackage[T1]{fontenc} 
\usepackage[english]{babel} 
\usepackage{amsmath,amssymb} 
\usepackage{graphicx} 
\usepackage{multirow} 
\usepackage[usenames,dvipsnames]{color} 

\usepackage[breaklinks,colorlinks,linkcolor=blue,citecolor=blue,urlcolor=blue]{hyperref}

\newcommand{\eqr}[1]{Eq.~\eqref{#1}}

\newcommand{\pd}[1]{\partial_{#1}}
\newcommand{\mt}[1]{\mathrm{#1}}
\newcommand{\expec}[1]{\langle\!\langle #1 \rangle\!\rangle}
\newcommand{\Pth}{P_{\mt{th}}}

\newcommand{\im}[0]{\mathrm{i}}

\newcommand{\Ln}[0]{\mathcal{L}_{\mathrm{n}}}
\newcommand{\Lphi}[0]{\mathcal{L}_{\phi}}
\newcommand{\Ls}[0]{\mathcal{L}_{s}}
\newcommand{\Lc}[0]{L_{c}}

\newcommand{\meV}[0]{\mt{m} \mt{e} \mt{V}}

\begin{document}

\title{Cavity Polariton Condensate in a Disordered Environment}

\author{Martin Thunert}
\altaffiliation{Both authors contributed equally to this work.}
\affiliation{Institut f\"ur Experimentelle Physik II, Universit\"at Leipzig, 04103 Leipzig, Germany}

\author{Alexander Janot}
\altaffiliation{Both authors contributed equally to this work.}
\affiliation{Institut f\"ur Theoretische Physik, Universit\"at Leipzig, 04103 Leipzig, Germany}

\author{Helena Franke}
\affiliation{Institut f\"ur Experimentelle Physik II, Universit\"at Leipzig, 04103 Leipzig, Germany}

\author{Chris Sturm}
\affiliation{Institut f\"ur Experimentelle Physik II, Universit\"at Leipzig, 04103 Leipzig, Germany}

\author{Tom Michalsky}
\affiliation{Institut f\"ur Experimentelle Physik II, Universit\"at Leipzig, 04103 Leipzig, Germany}

\author{Mar\'{i}a Dolores Mart\'{i}n}
\affiliation{Departamento de F\'{i}sica de Materiales, Universidad Aut\'{o}noma de Madrid, Madrid 28049, Spain}
\affiliation{Instituto de Ciencia de Materiales ``Nicol\'{a}s Cabrera,'' \\ Universidad Aut\'{o}noma de Madrid, Madrid 28049, Spain}

\author{Luis Vi\~{n}a}
\affiliation{Departamento de F\'{i}sica de Materiales, Universidad Aut\'{o}noma de Madrid, Madrid 28049, Spain}
\affiliation{Instituto de Ciencia de Materiales ``Nicol\'{a}s Cabrera,'' \\ Universidad Aut\'{o}noma de Madrid, Madrid 28049, Spain}
\affiliation{Instituto de F\'{i}sica de la Materia Condensada, Universidad Aut\'{o}noma de Madrid, Madrid 28049, Spain}

\author{Bernd Rosenow}
\affiliation{Institut f\"ur Theoretische Physik, Universit\"at Leipzig, 04103 Leipzig, Germany}

\author{Marius Grundmann}
\affiliation{Institut f\"ur Experimentelle Physik II, Universit\"at Leipzig, 04103 Leipzig, Germany}

\author{R\"udiger Schmidt-Grund}
\affiliation{Institut f\"ur Experimentelle Physik II, Universit\"at Leipzig, 04103 Leipzig, Germany}

\date{\today}

\begin{abstract}
We report on the influence of disorder on an exciton-polariton condensate in a ZnO based bulk planar microcavity and compare experimental results with a theoretical model for a non-equilibrium condensate.
Experimentally, we detect intensity fluctuations within the far-field emission pattern even at high condensate densities which indicates a significant impact of disorder.
We show that these effects rely on the driven dissipative nature of the condensate and argue that they can be accounted for by spatial phase inhomogeneities induced by disorder, which occur even for increasing condensate densities realized in the regime of high excitation power.
Thus, non-equilibrium effects strongly suppress the stabilization of the condensate against disorder, contrarily to what is expected for equilibrium condensates in the high density limit.
Numerical simulations based on our theoretical model reproduce the experimental data.
\end{abstract}
\maketitle
%

\section{Introduction}

The observation of a macroscopically coherent quantum state of exciton-polaritons, a so-called polariton Bose-Einstein condensate (BEC),~\cite{Kasprzak.2006,Balili.2007} has opened an active and challenging research field. Exciton-polaritons (for brevity polaritons) are mixed light-matter excitations in a microcavity (MC).~\cite{Carusotto.2013,Deng.2010} At finite quasi-particle density, several fascinating phenomena like superfluidity~\cite{Amo.2009,Amo.2009b,Sanvitto.2010} and the formation of quantum vortices~\cite{Lagoudakis.2008}, were discovered. This allows for numerous novel applications like optical parametric oscillators~\cite{Baumberg.2000}, polariton lasers~\cite{Schneider.2013,Bhattacharya.2014} and logical elements~\cite{Bajoni.2008,Steger.2012,Ballarini.2013,Anton.2013b,Sturm.2014}, which are usually restricted to low temperatures. However, polariton BECs even at room-temperature were observed in MCs based on wide band gap materials like GaN~\cite{Christopoulos.2007,Christmann.2008,Daskalakis.2013} and ZnO~\cite{Lu.2012,Li.2013,Lai.2012} or organic materials~\cite{Plumhof.2014}, paving the way for technological applications.
At the moment, experiments in these materials are significantly affected by disorder~\cite{Christopoulos.2007,Franke.2012,Trichet.2013}, and a thorough understanding of the impact disorder has on experimental observables in a polariton BEC is called for.

In contrast to conventional BECs, occurring for example in cold atom systems, polaritons have a finite lifetime, which gives rise to unique properties of the condensate. Nonetheless, there remain similarities, for instance, in the absence of disorder quasi-long range order of a two-dimensional polariton condensate~\cite{Roumpos.2012,Chiocchetta.2013,Spano.2012,Spano.2013} and superfluidity is theoretically expected~\cite{Wouters.2010,Keeling.2011} and experimentally observed.~\cite{Amo.2009,Amo.2009b,Sanvitto.2010}
However, recent theoretical studies have revealed exciting differences between equilibrium and non-equilibrium condensates~\cite{Sieberer.2013,Sieberer.2014,Tauber.2014,Altman.2013,Janot.2013}. 
For example, it is predicted that correlation functions for the condensate wave function decay exponentially~\cite{Altman.2013} and that superfluidity vanishes in the presence of disorder.~\cite{Janot.2013}

A polariton BEC is a steady state out of equilibrium where losses are compensated by external excitation. 
In the presence of disorder, spatial inhomogeneities of the condensate phase are induced.~\cite{Janot.2013} If the phase fluctuates on length scales comparable to the condensate size, spatial correlations and phase rigidity are strongly reduced. In our work we will show that this leads to significant traces of disorder in the experimentally observed $k$-space intensity distribution, and theoretically demonstrate that the ratio of the condensate correlation length to the condensate size is independent of the condensate density. Consequently, in polariton condensates the stabilization against disorder fluctuations with increasing condensate density is strongly suppressed as compared to condensates in equilibrium.

This prediction is supported by experimental investigations of the impact of disorder on a two-dimensional polariton BEC in a ZnO based MC. We measure the $k$-space intensity distribution as a function of excitation power, or rather condensate density, and observe significant disorder effects even at high densities. 
Numerical simulations allow to compare with experimental data confirming our theoretical predictions.

For an equilibrium BEC our observations would be unexpected, since an increasing density screens the disorder potential and leads to an ordered superfluid state~\cite{Nattermann.2008,Falco.2009,Malpuech.2007}. Analogously, for a polariton BEC, interactions also can lead to superfluidity, as observed in clean samples~\cite{Amo.2009,Amo.2009b,Sanvitto.2010}. However, as mentioned above, in the presence of disorder the polariton BEC is strictly speaking not a superfluid and long-range order is destroyed.~\cite{Janot.2013} Thus, we expect and observe that disorder affects a dissipative polariton BEC much more than an equilibrium one.
Several further observations found in literature seem to support this. For example, in one-dimensional CdTe MCs~\cite{Manni.2011,Stepnicki.2013} and ZnO MCs~\cite{Trichet.2013} the spatial first-order correlation function of polariton BEC emission was analyzed in the presence of disorder and significant changes due to disorder were found. In the CdTe MCs the disorder effects remain present even with increasing excitation power, similarly to our findings in two-dimensional ZnO MCs. We note that the correlation length of the assumed disorder potential discussed in Ref.~\onlinecite{,Stepnicki.2013} is of the order of microns, which enables the trapping of the entire condensate. This is explicitly excluded in our model, since the disorder correlation length is assumed to be much smaller than the condensate size leading to spatial density and phase fluctuations of the condensate instead. 
Moreover, in various works on two-dimensional polariton BECs in CdTe based MCs disorder effects were also observed, leading to fluctuations within the far-field photoluminescence (PL) distribution~\cite{Richard.2005} or the spatial first-order correlation function~\cite{Kasprzak.2006}. Even frequency desynchronization between spatially separated condensate fragments can be induced, if the ratio between the disorder potential and the polariton interaction potential strength exceeds a critical value.~\cite{Baas.2008,Krizhanovskii.2009,Wouters.2008b,Eastham.2008} However, the dependence of the condensate density on the disorder effects was not analyzed within these works.

The paper is organized as follows: In Sec.~\ref{sec:theory_disorder} we introduce our theoretical model. We discuss the disorder impact on a homogeneously and inhomogeneously excited condensate for a quasi-equilibrium (weak gain and loss) and driven dissipative (strong gain and loss) condensate, respectively. Furthermore, we provide a general argument that explains our experimental findings. These are presented in Sec.~\ref{sec:experimental_results}. In Sec.~\ref{sec:fit_of_exp_data} the theoretical predictions are confirmed by comparing experimental data to theoretical simulations. 
The summary and conclusion can be found in Sec.~\ref{sec:conclusion}.


\section{Theoretical Predictions}
\label{sec:theory_disorder}
\subsection{Model}
\label{sec:theory_model}
%
A phenomenological description of the dynamics of the polariton condensate wave function~$\Psi(\vec{x},t)$ is given by an extended Gross Pitaevskii equation (eGPE)~\cite{Wouters.2007,Keeling.2008}  
\begin{align}
	\label{eq:eGPE}
	\im \hbar \pd{t} \Psi = &\left( -\frac{\hbar^2}{2 m} \vec{\nabla}^2 + V(\vec{x}) + U \left|\Psi\right|^2 \right)\Psi \\
	\nonumber
		&+ \im \left(R(\vec{x}) - \Gamma \left|\Psi\right|^2\right) \Psi \ ,
\end{align}
where $m$ is the effective mass of the lower polariton branch, $V$ an external potential and $U>0$ an onsite interaction constant.
The function~$R(\vec{x})$ describes the linear part of gain and loss due to inscattering from a reservoir of non-condensed polaritons and the finite lifetime of the condensate. The non-linearity $\Gamma |\Psi|^2$ implements a density dependent gain saturation with $\Gamma$ as gain depletion constant. 
 Since the propagation of the reservoir polaritons can be neglected, the spatial shape of $R(\vec{x})$ can be related to the Gaussian profile of the excitation laser, namely
\begin{align}
	\label{eq:reservoirfct}
	R(\vec{x}) = \hbar \gamma_{\mt{c}} \left(\frac{P}{P_{\mt{th}}} e^{-\vec{x}^2/\xi_P^2} - 1 \right) \ .
\end{align}
The parameter $\gamma_{\mt{c}}$ is the condensate decay rate (inverse lifetime $\gamma_{\mt{c}} = 1/\tau$). The ratio $P/P_{\mt{th}}$ is the excitation power versus its value at threshold~$\Pth$ at which condensation is observed first, and $\xi_P$ is the waist size of the Gaussian pump spot.
We note that for the case of a spatially homogeneous excitation the eGPE~\eqref{eq:eGPE} was successfully used to analyze a driven dissipative condensate.~\cite{Sieberer.2013,Altman.2013}

Because of interactions, the condensate energy is blueshifted by $n_0U$ where $n_0$ is the mean condensate density determined by the balance of gain and loss (for a definition of $n_0$ see \eqr{eq:defn0}). The healing length $\xi \equiv \hbar / \sqrt{2 m n_0 U}$ is obtained by comparing kinetic and interaction energy of \eqr{eq:eGPE}.

The disordered environment is described by a random potential~$V(\vec{x})$. We choose Gaussian-distributed delta-correlated disorder with zero mean and variance~$\xi_V^2 V_0^2$, see Appendix~\ref{sec:theory_detailsmodel} for details. We introduce an effective dimensionless disorder parameter,
\begin{align}
  \label{eq:kappa}
  \kappa \equiv \frac{\xi_V \: V_0}{\xi\: n_0U} \ .
\end{align}

An analysis of the gain and loss terms in \eqr{eq:eGPE} allows us to define a 'non-equilibrium parameter'
\begin{align}
  \label{eq:alpha}
  \alpha \equiv \frac{\Gamma}{U} \ .
\end{align}
Its magnitude parametrizes the influence of gain and loss on the polariton BEC. For example, in the limit $\alpha \to 0$ (keeping $n_0$ finite) the equilibrium mean field description of a BEC is obtained, and, on the other hand, in the limit $\alpha \to \infty$ the condensate is totally dominated by gain and loss. 

In this work, we will focus on single-mode steady-state solutions and therefor make the ansatz $\Psi(\vec{x},t) = \Psi(\vec{x}) \exp( - \im \omega t)$, where $\hbar \omega$ is the condensate energy. However, in experimental realizations more than one condensate mode can exist.
For any further details we refer to Appendix~\ref{sec:theory_detailsmodel}.

\subsection{Disorder Effects}
%
%
\subsubsection{Infinite condensate size}
Before we discuss a finite size polariton BEC we would like to consider a homogeneously excited condensate ($\xi_P \to \infty$), such that the reservoir function~\eqr{eq:reservoirfct} is a constant in space. We will i) review disorder effects on an equilibrium condensate~\cite{Nattermann.2008,Falco.2009}, and ii) describe differences to a polariton BEC (driven dissipative condensate)~\cite{Janot.2013}. 

\emph{Equilibrium condensate i):}
The disorder potential attempts to pin the condensate into its minima, whereby the energy costs for density deformations (kinetic term in \eqr{eq:eGPE}) have to be compensated. The balance of pinning and kinetic energy determines the density Larkin length $\Ln \approx \sqrt{\pi} \: \hbar^2 / m \: \xi_V V_0$~\cite{ImryMa.1975,Nattermann.2008,Falco.2009}. 
On the other hand, for a sufficiently large interaction energy~$n_0U$ the disorder gets screened.~\cite{Falco.2009}
The ratio of healing to Larkin length, $\xi/\Ln \sim \kappa$, describes this competition of disorder and interaction. For $\xi \ll \Ln$ ($\xi \gg \Ln$) the interaction energy is large (small) as compared to the disorder potential.  Due to the fact that the interaction energy increases with increasing density (and $\xi \propto 1/\sqrt{n_0}$), $\xi/\Ln$ decreases with increasing density, and disorder effects will fade away in this limit. Thus, for sufficiently high densities an equilibrium condensate will be ordered and superfluid.~\cite{Falco.2009}

%

\emph{Non-equilibrium condensate ii):}
In a driven system the mean density~$n_0$ of the condensate is determined by a balance of gain and loss. Disorder induces density fluctuations about this mean value. In a region with reduced density, as compared to~$n_0$, the gain mechanism tries to compensate the depletion, and more particles are scattered into the condensate than decay. On the other hand, in a region with increased density more particles decay than are injected from the reservoir. By virtue of the continuity equation, these local particle sources and sinks are connected by condensate currents. Because the density fluctuates randomly in space, a random distribution of sources and sinks forms and, thus, a random pattern of current flow is generated. The condensate current is proportional to the product of the density and the gradient of the condensate phase. Since the current is not constant, the phase cannot vary uniformly in space, and thus a random current configuration gives rise to a spatially fluctuating phase.
We note that in this work the term 'fluctuations' will be used for random spatial inhomogeneities.
The correlation length, over which the phase typically varies by $2\pi$, is given by $\Lphi \approx \sqrt{2}\pi \, \Ln / \alpha$. This scale can be obtained by a generalized Imry-Ma argument~\cite{Janot.2013}: a condensate current flowing out of (or into) a region of diameter $\Lphi$ is generated by an effective source (or sink) determined through an area average of multiple random sources and sinks. In contrast to an equilibrium condensate ($\alpha \to 0$ with $\Lphi \to \infty$), the phase fluctuations occurring in the case~$\Lphi < \infty$ destroy the quasi-long-range order of the condensate. As a consequence of these phase fluctuations, the superfluid stiffness vanishes in the thermodynamic limit even for weak disorder, and a superfluid behavior is only present below a finite length scale, namely the superfluid depletion length $\Ls \approx \sqrt{2} \pi \, \Ln / \alpha^2$.~\cite{Janot.2013}

\subsubsection{Finite condensate size}
From the analysis above we conclude that in a disordered environment a condensate of size $\Lc \ll \Lphi$ will behave completely different from one of size $\Lc \sim \Lphi$.
%
%
\begin{figure}[tb]
 	\centering
	\includegraphics[width = 1\hsize]{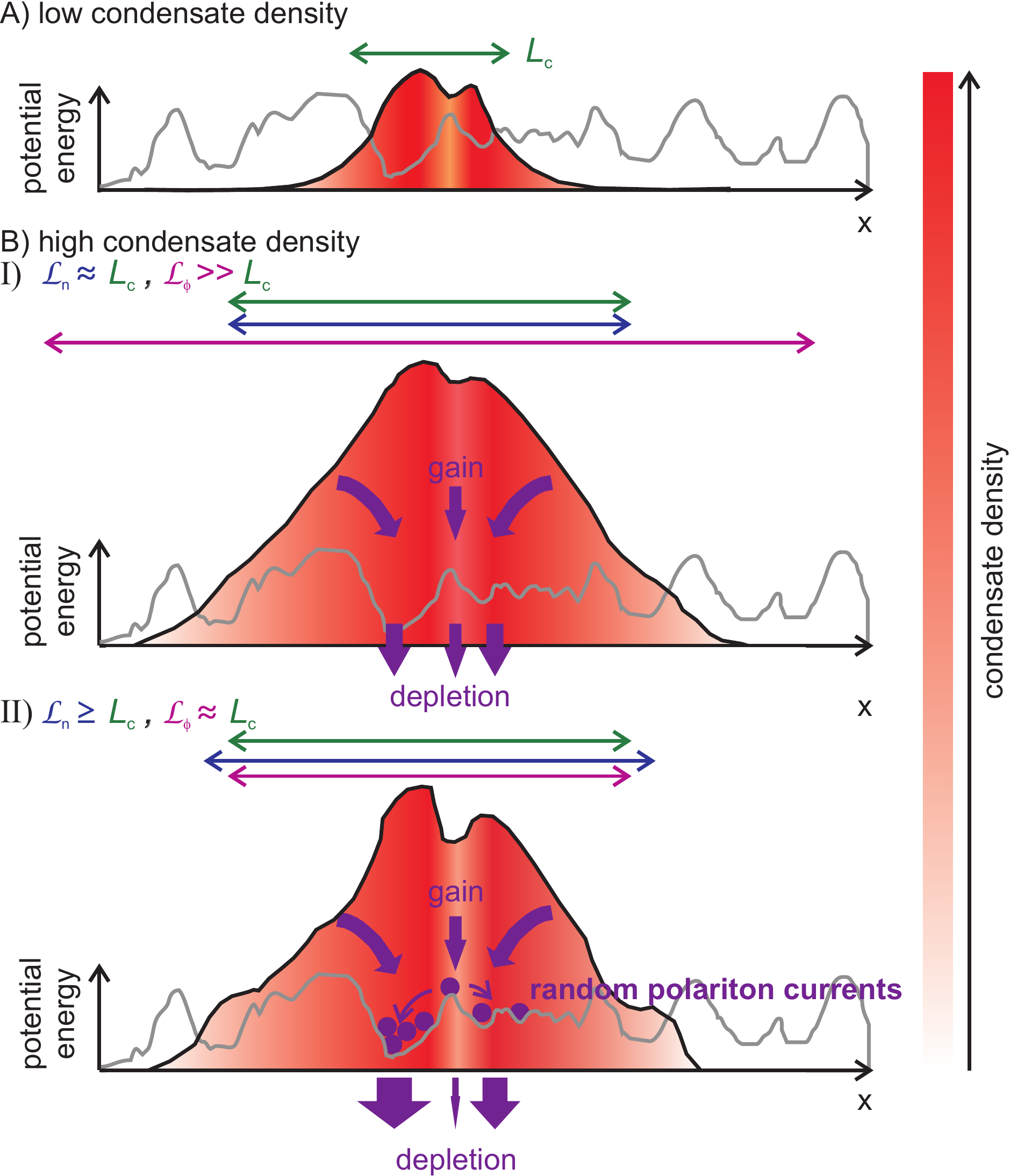}
 	\caption{(color online)
	Disorder impact on a polariton condensate for increasing density. The gray lines represent the disorder potential. The condensate density is depicted by a red color code and the corresponding interaction potential by a black line. In all cases, a Gaussian-shaped excitation spot is assumed. 
For low condensate densities (case A) significant disorder effects are present. 
Depending on the non-equilibrium nature of the condensate two different scenarios are expected for a high density (case B).
Scenario I (quasi-equilibrium condensate): disorder induces density fluctuations ($\Ln \sim \Lc$, $\Lc$ - condensate size), whereas the condensate phase remains unperturbed ($\Lphi \gg \Lc$). For sufficiently high densities the interaction potential screens the disorder, which results in a weakly perturbed condensate.
Scenario II (driven dissipative condensate): the presence of disorder in combination with gain and loss leads to phase fluctuations ($\Lphi \sim \Lc$). These are density independent, and thus disorder effects persist with increasing density.
}
	\label{fig:sketch_scenarios}
\end{figure}
In the following, we discuss these two scenarios sketched schematically in Fig.~\ref{fig:sketch_scenarios}B.

For scenario~I with $\Lc \ll \Lphi$ (called quasi-equilibrium in the following) the phase is correlated over the entire condensate region, and disorder induces mainly density fluctuations. As discussed above, the impact of disorder will decrease with increasing density, which should be directly observable by increasing the excitation power.
Such kind of percolation transition from a disordered to an ordered regime was predicted (for a polariton BEC in equilibrium) in Ref.~\onlinecite{Malpuech.2007}.

In the presence of gain and loss disorder induces phase fluctuations as explained above. For scenario~II we assume that the phase correlation length~$\Lphi$ is comparable to the condensate size~$\Lc$, i.e. $\Lphi \sim \Lc$, such that spatial correlations and superfluidity are destroyed.
The ratio $\Lc/\Lphi \propto (V_0 \ \xi_{\mt{P}} \xi_{\mt{V}} m / \hbar^2) \: (\Gamma / U)$ does not depend on the condensate density and, thus, is independent of the excitation power. A similar conclusion holds for the ratio~$\Lc / \Ls$. As a consequence, a condensate stabilization with increasing density, as present in an equilibrium system, is strongly suppressed.

%
\begin{figure}[tb]
	\centering
	\includegraphics[width=1\hsize]{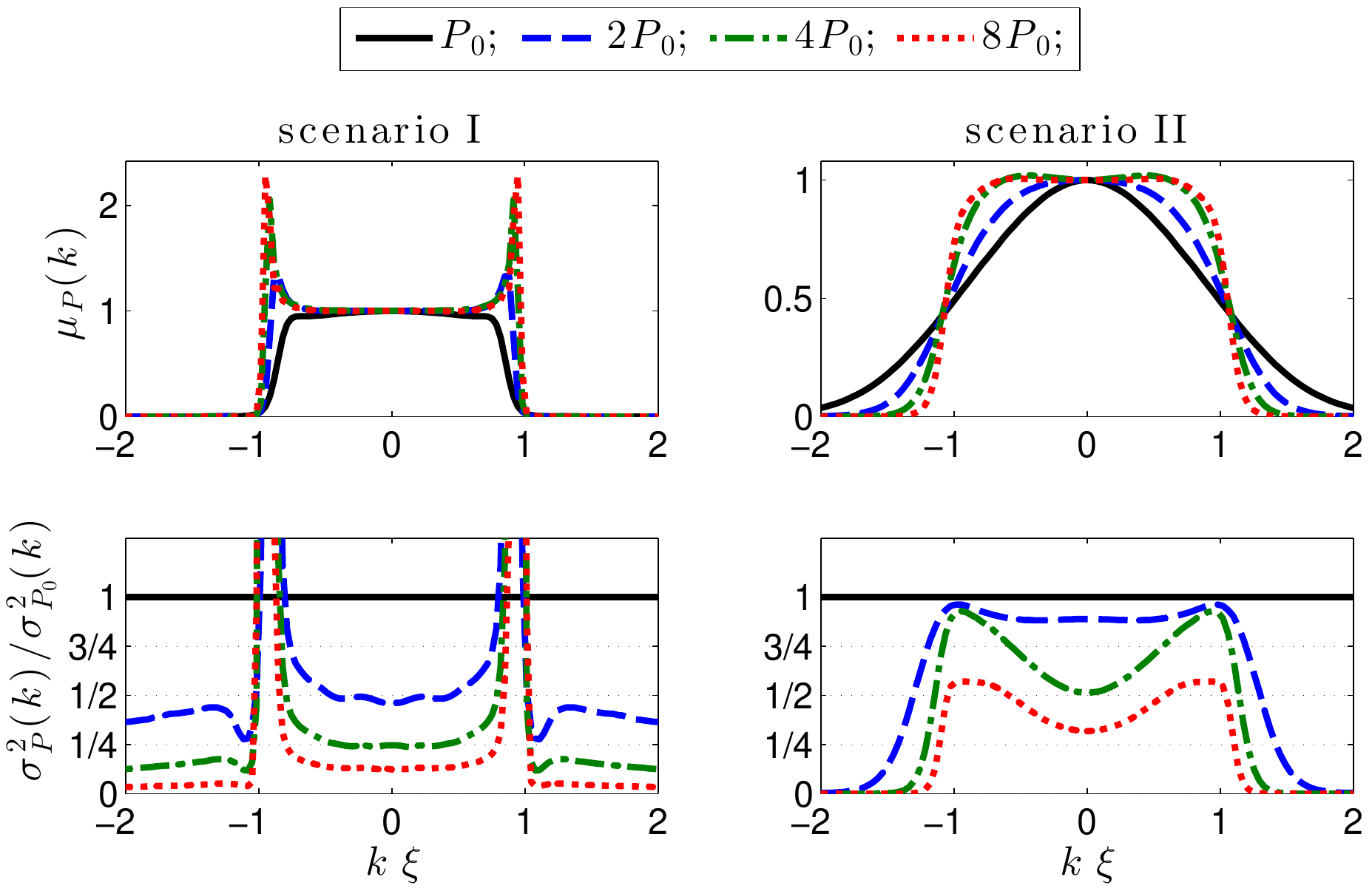}
	\caption{(color online)
Expectation value $\mu_P$ (upper row) and variance $\sigma^2_P$ (lower row) of the (normalized) intensity distribution $I_P(k)$. In order to compare the fluctuations for increasing excitation powers we present the ratio $\sigma^2_P(k) / \sigma^2_{P_0}(k)$ with $P_0 = 4 P_{\mt{th}}$.
The left and right column depict a quasi-equilibrium (scenario I) and a driven dissipative (scenario II) condensate, respectively. For wavevectors $|k \xi| \lesssim 1$, scenario I shows a linear reduction of fluctuations with inverse excitation power, 
$\sigma^2_P(k) \propto 1/P$,  while scenario II exhibits a suppressed stabilization with increasing excitation power.
We averaged $1560$ disorder realizations and used: $\Ln/\Lc = 1 \: (4.5)$, $\Lphi/\Lc = 10 \: (2.5)$, $\Ls/\Lc = 20 \: (0.3)$ for scenario~I (II).
}
	\label{fig:IkMeanVar}
\end{figure}
In order to make our analysis more quantitative, we have studied theoretically the excitation power dependence of the two-dimensional $k$-space intensity $I_P({\vec{k}}) \propto \gamma_c |\Psi_{\vec{k}}|^2$ which can be directly compared to experimental data. To this end, \eqr{eq:eGPE} was simulated for many disorder realizations (see Appendix~\ref{sec:theory_numsim} for details). We have extracted the expectation value, denoted by $\mu_P(k)$, and the variance, denoted by $\sigma^2_P(k)$, of the normalized intensity $I_P(\vec{k})$ by averaging over disorder configurations. We note that for a sufficiently large number of realizations, the disorder average restores radial symmetry, such that  the expectation values  $\mu_P(k)$ and $\sigma^2_P(k)$  depend on the magnitude $k = |\vec{k}|$ of wavevector only.

In Fig.~\ref{fig:IkMeanVar}, the results for $\mu_P$ and $\sigma_P^2$ are shown for scenario I (left panels) and II (right panels). We find that the intensity $I_P$ vanishes for all wavevectors outside of the lower-polariton dispersion ( $k \gtrsim \xi^{-1}$) and that its average value does not change qualitatively as compared to a disorder-free system (cf.~Ref.~\onlinecite{Wouters.2008}). However, for a single snap-shot (see Fig.~\ref{fig:snapshot}) disorder breaks the radial symmetry and induces intensity fluctuations proportional to~$\sigma_P$.
For scenario I and for wavevectors $|k| \lesssim \xi^{-1}$, these fluctuations decay linearly with inverse excitation power, in agreement with the expectation $\sigma^2_P \sim \kappa^2 \propto 1/P$ for $\kappa \ll 1$.
We note  that  regions with $k \approx \xi^{-1}$ show a high ratio $\sigma_P / \sigma_{\! P_{0}}$ (peaks in Fig.~\ref{fig:IkMeanVar} lower left panel).
In this $k$-region, the emission intensity is increasing very rapidly with excitation power (see Fig.~\ref{fig:IkMeanVar} upper left panel), because of the repulsive potential hill created by the finite excitation spot.~\cite{Wouters.2008} Thus, the increase of  fluctuation strengths with excitation power for $k \approx \xi^{-1}$ is really due to the increase of emission power and does not yield information about the screening of the disorder potential for high condensate densities. 
 
For scenario~II, the stabilization with increasing excitation power is suppressed (see lower right panel of Fig.~\ref{fig:IkMeanVar}). As compared to scenario~I, the decrease of $\sigma^2_P$ with increasing condensate density is weaker than $\sigma^2_P \propto 1/P$. These findings agree well with our argument provided above.

The reservoir of non-condensed polaritons interacts with the condensate and thus leads to an increase of the blueshift.~\cite{Wouters.2007,Wertz.2010,Anton.2013c} Usually, this is accounted for by adding a potential term proportional to the reservoir density in \eqr{eq:eGPE}.~\cite{Wouters.2007} Such a term will modify the emission frequency of the condensate (real part of \eqr{eq:eGPE}), however, does not change the non-equilibrium continuity equation (imaginary part of \eqr{eq:eGPE}). Hence, the mechanism of generating random condensate currents is not altered qualitatively by reservoir-condensate interaction and, thus, we believe that they can be safely neglected for our analysis.

\section{Experiment}
\label{sec:experimental_results}

In this section we discuss the experimentally observed behavior of the far-field PL emission pattern of a polariton condensate in a ZnO-based MC with pronounced structural disorder as a function of excitation power. For this experiment, the sample was excited using a pulsed Nd:YAG laser with a pulse duration of $500~\mt{ps}$. This is three orders of magnitude larger than the polariton relaxation time (0.4~$\mt{ps}$) which is determined from the spectral linewidth of the condensate emission. Thus, we can assume a quasi--continuous-wave excitation, which justifies the comparison with numerical simulations based on a steady state theory as will be discussed in Sec.~\ref{sec:fit_of_exp_data}. Further details about the experimental setup can be found in Appendix~\ref{sec:sample_setup}. The MC consists of a half wavelength ZnO cavity, which simultaneously acts as active medium, showing a quality factor of about 1000 and a maximum coupling strength of about 45~meV ($\Omega_{\mt{Rabi}} \approx 90~\meV$) at $T = 10~\mt{K}$. By using a wedge-shaped cavity, the detuning between the cavity mode energy and the excitonic transition energy strongly varies with the lateral sample position. Structural investigations (atomic force microscopy, X-ray diffraction, cross-sectional transmission electron microscopy) yield a smooth but polycrystalline cavity layer, exhibiting a low interface roughness of $R_{\mt{rms}}$~=~1.9~nm. Furthermore, the cavity layer is 	preferentially $c$-plane oriented and laterally textured, containing large grains aligned in the growth direction reaching from the bottom to the top (grain sizes ranging from 20~nm up to 120~nm). Further information about the sample properties can be found in Ref.~\onlinecite{Franke.2012}. 
Due to the textured structure we suppose that an electronic disorder potential is primarily caused by depletion of carriers, e.g. aluminum donor bound excitons~\footnote{aluminum is a common donor in ZnO layers and as a component of the used sapphire substrate and Bragg mirror layers it can easily diffuse into the ZnO cavity during the annealing process at high temperatures of $T~\approx~900~^\circ$}, due to interface band bending at grain boundaries~\cite{Orton.1980}. (see Appendix~\ref{sec:experiment_disorderorigin} for details).

\begin{figure}[tb]
	\centering
	\includegraphics[width=0.9\hsize]{./{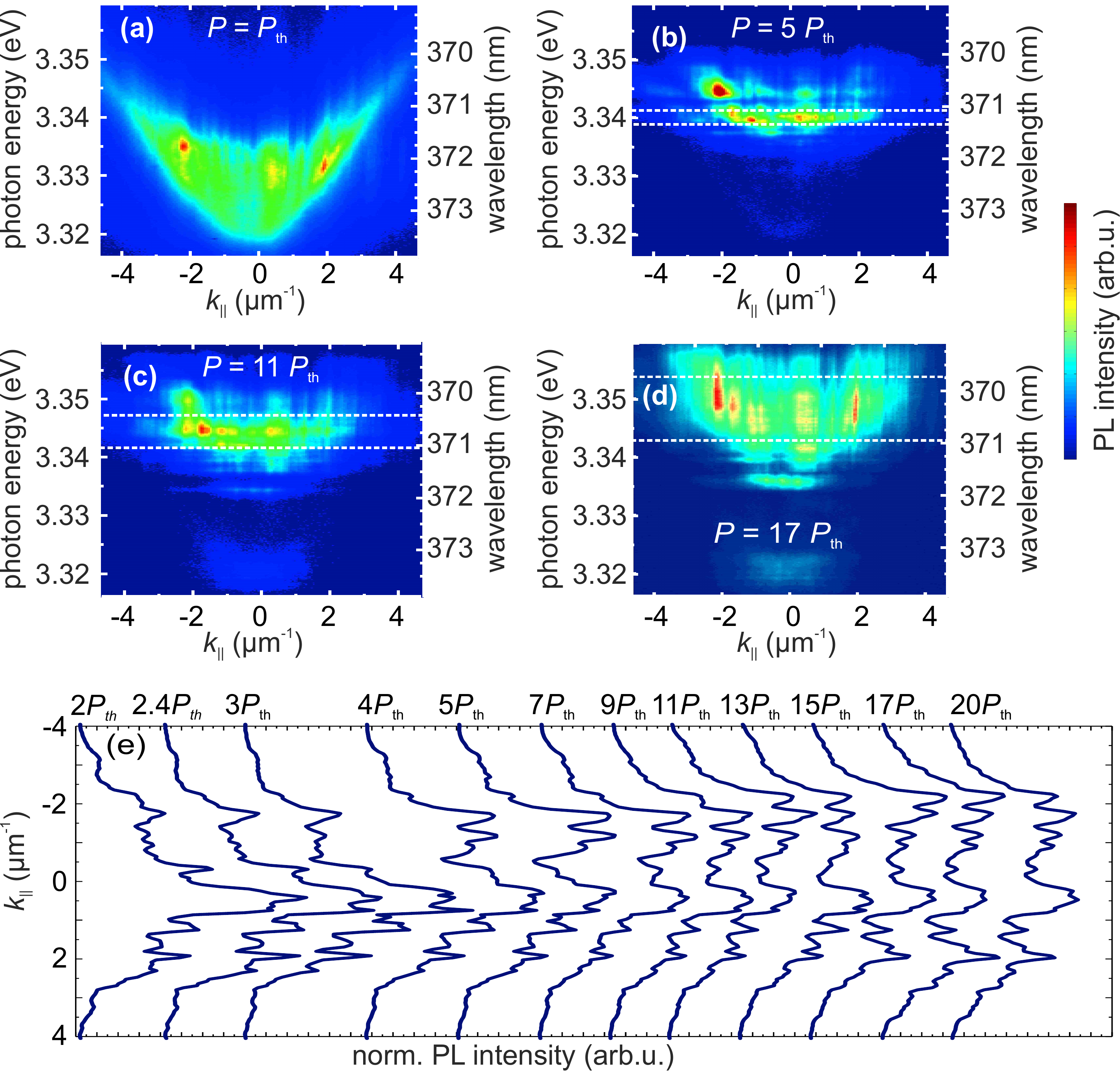}}
	\caption{(color online)
(a)-(d): Excitation power series of the far-field PL emission in a linear false color scale for $T = 10~\mt{K}$ and a detuning of $\Delta = -30~\meV$. The excitation power is normalized to the condensation threshold $P_{\mt{th}}$. (e) $I_P(k)$ profiles of the condensate are extracted. These are integrated over the energy range $\Delta E$ marked by the white lines in the far-field emission pattern. The PL intensity for each power is normalized to the mean value of each $I_P(k)$ profile.}
	\label{fig:Disorder}
\end{figure}
Figures~\ref{fig:Disorder}(a)-(d) show the excitation power dependence of the PL $k$-space emission pattern for $T = 10 \, \mt{K}$ and detuning $\Delta = -30 \, \meV$. We deduce a polariton effective mass of $m=4.4 \times 10^{-5}~m_{\mt{e}}$ ($m_{\mt{e}}$: free electron mass) from the dispersion of the lower polariton branch (LPB) (not shown here). The excitation power density at condensation threshold is $P_{\mt{th}} = 79~\mt{W}\mt{cm}^{-2}$. Note that the determination of the excitation power density at threshold is quite complex, e.g. due to the coexistence of intense emission from uncondensed polaritons for $P \gtrsim \Pth$, but significant for the comparison with theoretical calculations discussed in Sec.~\ref{sec:fit_of_exp_data}. Details for the experimental determination of $\Pth$ can be found in the Supplemental Material, Sec.~\ref{sec:SM_Pth}.

In all cases investigated here, the condensate emission is distributed dispersion-less at horizontal lines in $k$-space with maximum intensity between the LPB dispersion, which is visible in the far-field PL images (cf.~Fig.~\ref{fig:Disorder}) for low excitation power $P \gtrsim \Pth$. This indicates a weak expansion of the condensed polaritons due to the background potential induced by the excitation spot, whose size is similar or even larger than the polariton propagation length.~\cite{Franke.2012,Wouters.2008} For the lowest excitation power shown here, $P = P_{\mt{th}}$, the emission intensity from the uncondensed polaritons and the condensate are of same order which prevents a clear distinction.
With increasing excitation power the BEC states undergo a blueshift due to the increasing interaction potential,
and we observe several states with different energy.
Previous studies in the literature on this multimode behavior show that the emission from coexisting individual modes originates from different regions of the same condensate.~\cite{Krizhanovskii.2006,Baas.2008,Krizhanovskii.2009} However, other studies on polariton condensates in a disordered environment found that long-range spatial coherence is still present for their energy-averaged emission~\cite{Kasprzak.2006,Richard.2005} indicating persistent correlations between different, possibly spatially separated condensate states.

For a wide range of excitation powers, condensate emission out of two energy ranges is observed, which are stable and energetically well separated. For a further analysis we select only one of these energy channels, in order to compare with numeric simulations of a single-mode condensate, cf. Sec.~\ref{sec:fit_of_exp_data}.
In Fig.~\ref{fig:Disorder}(b)-(d) we marked the selected energy channel by two white dashed lines. This delimitation is defined by an energy range  $\Delta E$ which corresponds to the excitation power dependent full width at half maximum of the condensate emission. Fig.~\ref{fig:Disorder}(e) shows the far field emission profiles $I_P(k)$ for the selected energy channel and increasing excitation power, integrated over $\Delta E$.

The $I_P(k)$ profiles show several randomly distributed inhomogeneities and differ strongly from the smooth and ideally radial symmetric distribution expected for a disorder-free sample.~\cite{Wouters.2008} Remarkably, the intensity fluctuations persist even for high excitation power, i.e. high condensate densities.
We note that the constant sharp stripes in the $I_P(k)$ profile at a specified \textit{k} for all excitation powers are caused by imperfections of the setup, probably due to the microscope objective.

A similar finding with increasing excitation power was also observed for other detunings $\Delta = -50~\mt{meV},\ldots,-10~\mt{meV}$, and we conclude that our observation does not depend significantly on the particular choice of detuning within the mentioned range.

%
To investigate the temporal coherence properties of the condensate we used a Michelson interferometer in the plane mirror (PM) - retroreflector (RR) configuration to superimpose the PL emission of polaritons with opposite emission angles or rather wavevectors. For this experiment, the sample was excited by a frequency-tripled Ti:sapphire laser at 266~nm with a pulse duration of about 2~ps. Further details of the setup are provided in Appendix~\ref{sec:sample_setup}. 
\begin{figure}
	\centering
		\includegraphics[width=\linewidth]{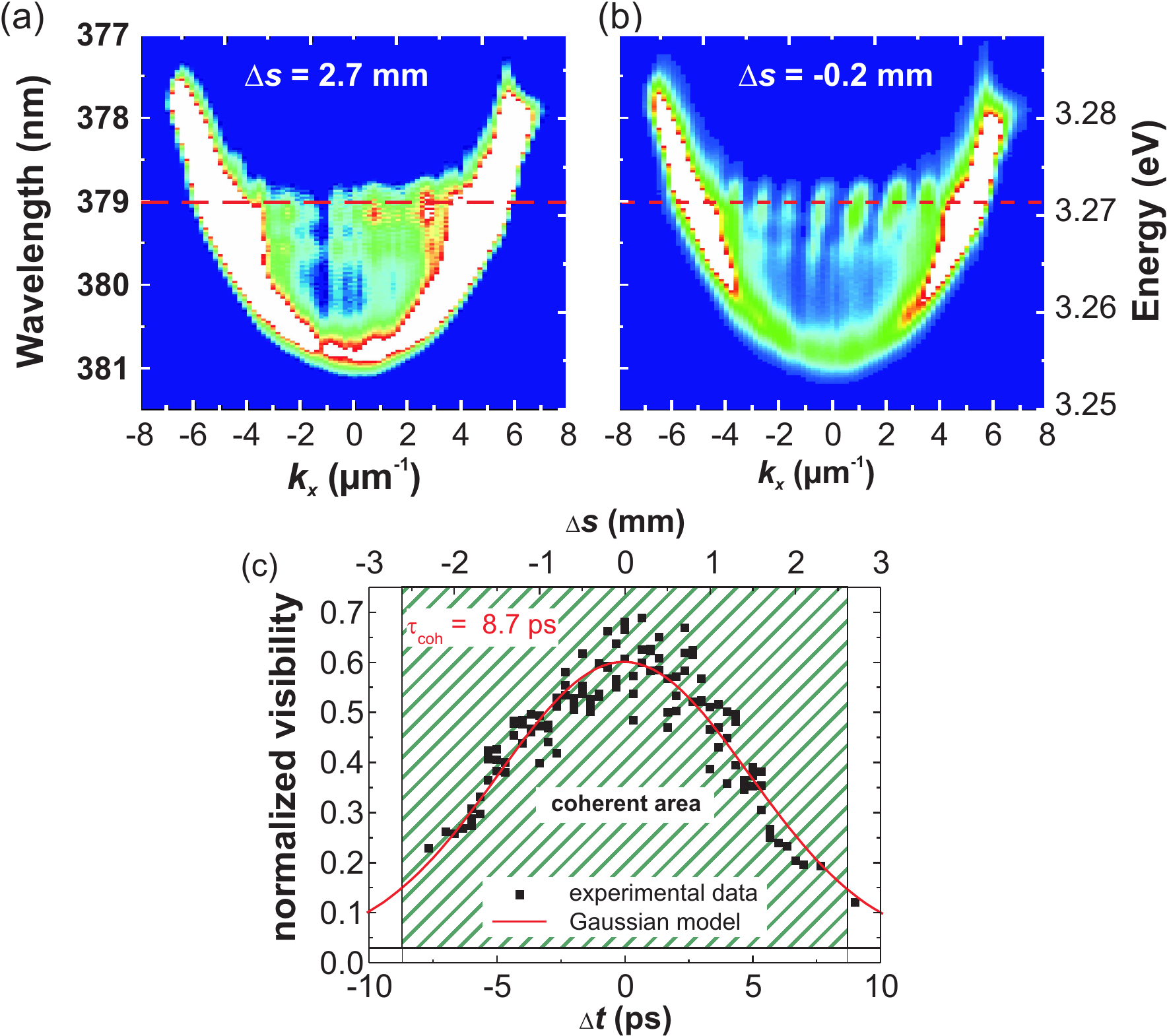}
	\caption{(color online)
(a)-(b): Energy-resolved far-field PL interference pattern in a linear false color scale for $T = 10~K$ and $\Delta = -47$~meV: (a) large path difference~$\Delta s$: uncorrelated far-field PL emission (basically the sum of the emission from the individual interferometer arms), (b) path difference close to zero: distinct interference fringes indicating mutual temporal coherence of the polariton emission. The red dashed line indicates the energy of the investigated condensate state. (c) Normalized visibility of the interference fringes as a function of the path difference.}
	\label{fig:EkSpace_Interference}
\end{figure}
The RR is mounted on a motorized linear stage, that allows us to vary the path difference~$\Delta s$ between the emission collected from both interferometer arms. Fig.~\ref{fig:EkSpace_Interference}(a)-(b) show two selected interferograms of the $I(E,k)$ emission pattern for large (Fig.~\ref{fig:EkSpace_Interference}(a)) and short (Fig.~\ref{fig:EkSpace_Interference}(b)) path differences~$\Delta s$, respectively. To investigate the temporal coherence properties of the polariton condensate we analyzed the normalized visibility of the interference fringes,
\begin{align}
	\label{eq:Inorm}
	V_{\mt{norm}} & = \frac{I_ {\text{interf}} - I_{RR} - I_{PM}}{2\sqrt{I_{RR}  I_{PM}}} = g^1(\Delta t) \cos(\phi_{12}),
\end{align}
as a function of the temporal delay $\Delta t = \frac{\Delta s}{c}$, where $c$ is the speed of light (cf. Fig.~\ref{fig:EkSpace_Interference}(c)). Here, $I_{\mt{interf}}$ is the intensity of the interference pattern, and $I_{\mt{RR}}$, $I_{\mt{PM}}$ are the intensities of the RR and PM arm, respectively, $g^1$ is the first-order coherence function and $\phi_{12}$ is the phase difference between the emission from the individual interferometer arms. By assuming a Gaussian decay of $g^1(\Delta t) = g^1(0) \exp[
- (\pi / 2) (\Delta t ^2 / \tau_{\mt{coh}}^2)]$\cite{Saleh.2008} we determined a coherence time of about $\tau_{\mt{coh} } = 8.7 \, \mt{ps}$. This is more than 50~times larger than the lifetime of the uncondensed polaritons of about 160~fs, which is deduced from the spectral linewidth of the polariton emission for $P \leq P_{\mt{th}}$ and for an energy range similar to the condensate energy at $P = P_{\mt{th}}$. Consequently, the coherence of the investigated quantum system is conserved during the multiple reabsorption and reemission processes, which can thus be identified as a condensate.
We note that the experimentally estimated coherence time is a lower limit for the real value. We identify two experimental artifacts that restrict the determination of the real condensate's coherence time, namely a spectrally and path difference dependent phase shift $\phi_{12}(\lambda, \Delta s)$ (artifact A) as well as a fast decay of the condensate emission intensity due to the short excitation pulses of about 2~ps that are used for the coherence time measurement (artifact B). By analyzing the impact of these artifacts quantitatively (cf. Supplemental Material, Sec.~\ref{sec:SM_limits_cohTime}), we roughly estimated the expected real values for the coherence time of $\tau^A_\mt{corr} = 10.3$~ps and $\tau^B_\mt{corr} = 14$~ps. By applying both corrections simultaneously, a maximum coherence time of $\tau_\mt{corr}$~=~24~ps was estimated.

For an ideal (homogeneous, disorder-free) condensate a linewidth of $\Delta E \approx 0.66 \,  h / \tau_\mt{coh} = 0.66 \, h / 8.7~\mt{ps} = 0.31~\mt{meV}$ would be expected for the condensate emission according to the Wiener-Khinchin theorem~\cite{Saleh.2008} (and even less assuming the corrected values for $\tau_\mt{coh}$), where $h$ is the Planck constant. This is about a factor of 6.5 smaller than the observed minimum linewidth of 2~meV for the condensate emission in this experiment. Since the investigated condensate is a complex quantum system including spatial density and phase fluctuations we assume that the Wiener-Khinchin theorem cannot be applied here. We rather suppose that the mechanism which causes a broadening of the emission linewidth (e.g. repulsive particle interaction \cite{Porras.2003}) does not affect the coherence time to the same extent. This is supported by the quantitative discrepancy between the emission linewidth and the coherence time, which is observed also in a CdTe \cite{Kasprzak.2006} as well as in a ZnO MC.~\cite{Lai.2012} We note that despite of the fast decay of polaritons, condensate emission can be observed up to 90 ps after the arrival of the exciting laser pulse, which thus allows for the experimental observation of coherence in the mentioned time range.

Summarizing, the experimental observations indicate a strong impact of disorder on the polariton BEC even at high excitation power well above the condensation threshold. As discussed in Sec.~\ref{sec:theory_disorder} the suppression of disorder effects with increasing condensate density is strongly hindered for a polariton BEC. We assume that the interplay of gain-loss and disorder prevents a stabilization at high excitation power also in the experiment.

\section{Comparison between Theoretical Model and Experiment}
\label{sec:fit_of_exp_data}
In the following, we will compare our experimental observations with numerical simulations. 

At threshold~$P=\Pth$ a cross-over from a non-condensed state to a polariton BEC takes place, typically indicated by a super-linear increase of the emission intensity. Such a transition is not very well described by the used eGPE~\eqref{eq:eGPE}. For this reason, the data analysis is done well above threshold, 
where both experimentally observed and theoretically calculated blueshift (condensate density) increase linearly with pump power.
We note that the evolution of the experimentally measured polariton blueshift~$\Delta E$ as a function of the excitation power shows two kinks at $P = 2~P_{\mt{th}}$ and $P = 4~P_{\mt{th}}$ (cf. Fig.~\ref{fig:Blueshift} in Appendix~\ref{sec:experiment_disorderorigin}). We believe that the slope of $\Delta E$ for $P < 2~P_{\mt{th}}$ is predominantly caused by an electronic disorder potential, which starts to saturate for $P = 2~P_{\mt{th}}$, and that for $P \geq 4~P_{\mt{th}}$ the blueshift is governed by condensate-condensate interactions. Further discussions are presented in Appendix~\ref{sec:experiment_disorderorigin} and references therein.
%

For the comparison between the theoretical model and the experimental data, the parameters of the eGPE~\eqref{eq:eGPE} are chosen according to the experiment, see Tab.~\ref{tab:parameters}. 
\footnote{The non-equilibrium parameter~$\alpha$ can be extracted by the linewidth divided by the derivative of the condensate blueshift w.r.t. excitation power, $\alpha = \hbar \gamma_{\mt{c}} / (d \Delta E / d(P/\Pth))$. Within the error-bounds we chose $\alpha = 7$ in order to reproduce the experimental data.
	}
We note that a quantitative determination of the disorder parameter from experiment is very challenging, cf. discussion in Appendix~\ref{sec:experiment_disorderorigin}, and we chose $\xi_{\mt{V}} V_0 \approx 0.15 \: \mu\mt{m} \, \meV$ for simulations.
%
\begin{table}[tb]
  \centering
  \begin{tabular}[c]{|c|c|c|c|c|c|c|c|c|}
	\hline
    $m$ & $\tau$ & $\xi_{\mt{P}}$ & $P_{\mt{th}}$ & $P/P_{\mt{th}}$ & $d \Delta \! E / d(P \! / \! \Pth)$ \\ 
	\hline
	$4.4 \cdot 10^{-5}\, \mt{m}_{\mt{e}}$ & $0.4 \, \mt{ps}$ & $2 \, \mu\mt{m}$ & $79 \, \mt{W} \! / \! \mt{cm}^2$ & $2,\ldots,20$ & $0.7 \: \mt{meV}$ \\ 
	\hline
  \end{tabular}
  \newline
  \begin{tabular}[c]{|c|c|c|c||c|c|c|}
	\hline
    $\alpha$ & $P/P_{\mt{th}}$ &  ${\xi_{\mt{V}} V_0}/{l_c \, \hbar \gamma_c}$ & ${\xi_{\mt{P}}}/{l_{\mt{c}}}$ & $\Ln / \Lc$   & $\Lphi / \Lc$  & $\Ls / \Lc$\\
	\hline
	$7$	& $5,11,15,20$ & $0.125$ & $3$ & $5$ & $3$ & $0.4$\\
	\hline
  \end{tabular}
\caption{Parameters extracted from experiment and corresponding parameters used for simulations as well as relevant length scales. For definitions see Appendix~\ref{sec:theory_detailsmodel}.}
\label{tab:parameters}
\end{table}

%
\begin{figure}[tb]
  \centering
	\includegraphics[width=0.95\hsize]{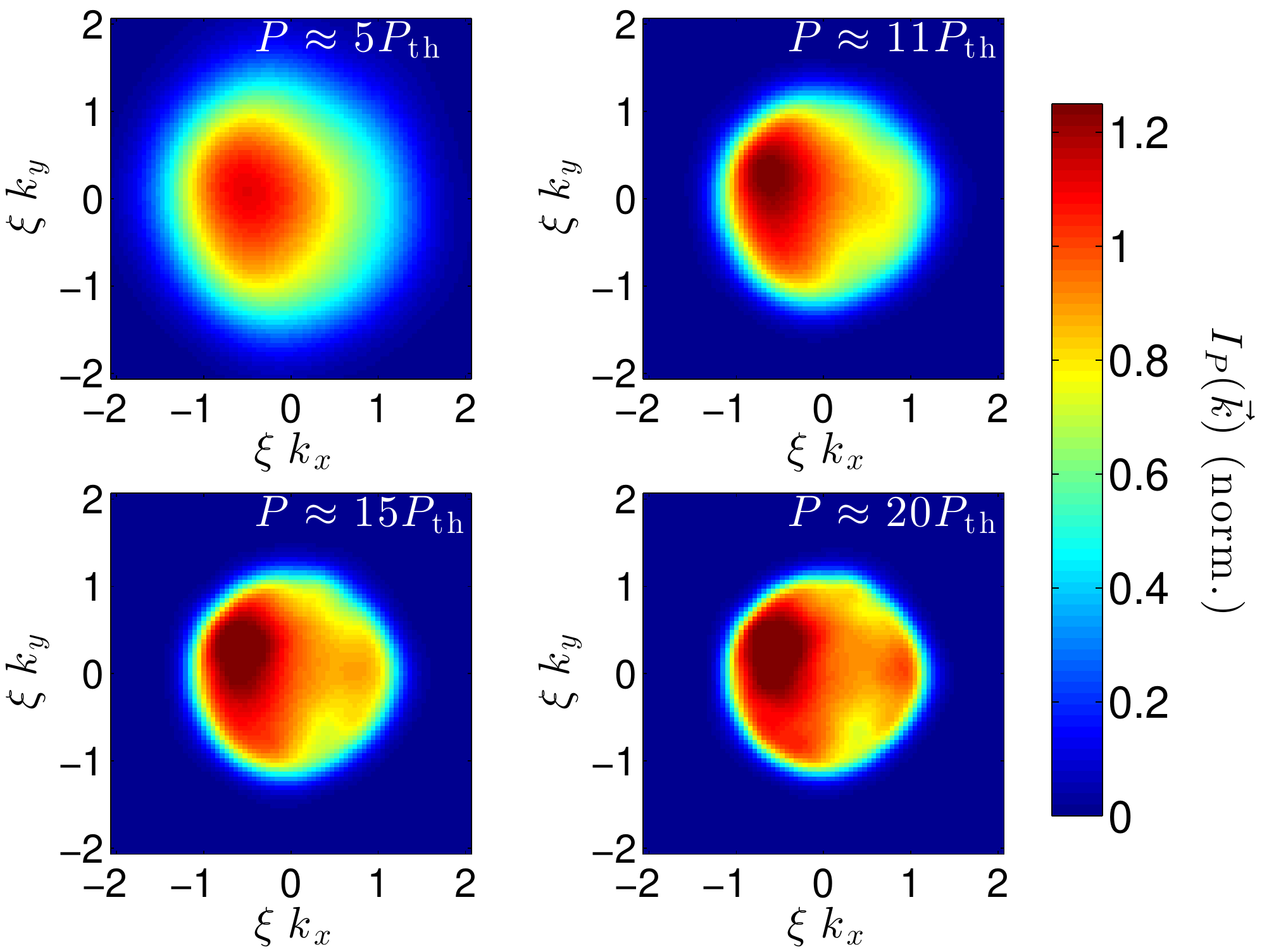}
  \caption{(color online)
Snap shots of the normalized two-dimensional intensity distribution~$I_P(\vec{k})$ of a polariton BEC with $\Lphi \sim \Lc$ in a disordered environment for increasing excitation power~$P$. The used parameters are presented in Tab.~\ref{tab:parameters}.
}
  \label{fig:snapshot}
\end{figure}
For a typical disorder realization, a series of numerically obtained snap shots of the two-dimensional intensity distribution $I_P(\vec{k})$ for increasing excitation power is shown in Fig.~\ref{fig:snapshot}. These images correspond to a polariton BEC described by scenario~II. We clearly observe a disorder-induced deviation from the ideally radial distribution, which does not converge to a symmetric intensity distribution while increasing the excitation power. Such an asymmetry as well as its persistence is also observed experimentally, see Fig.~\ref{fig:Disorder}, and thus agrees qualitatively with our simulations.
We note that the experimental data represent the intensity distribution of one disordered sample, and correspond to a one-dimensional cut along a given line crossing the origin of the two-dimensional $k$-space distribution, for example the $x$-axis.

\begin{figure}[tb]
	\centering
	\includegraphics[width=1.00\hsize]{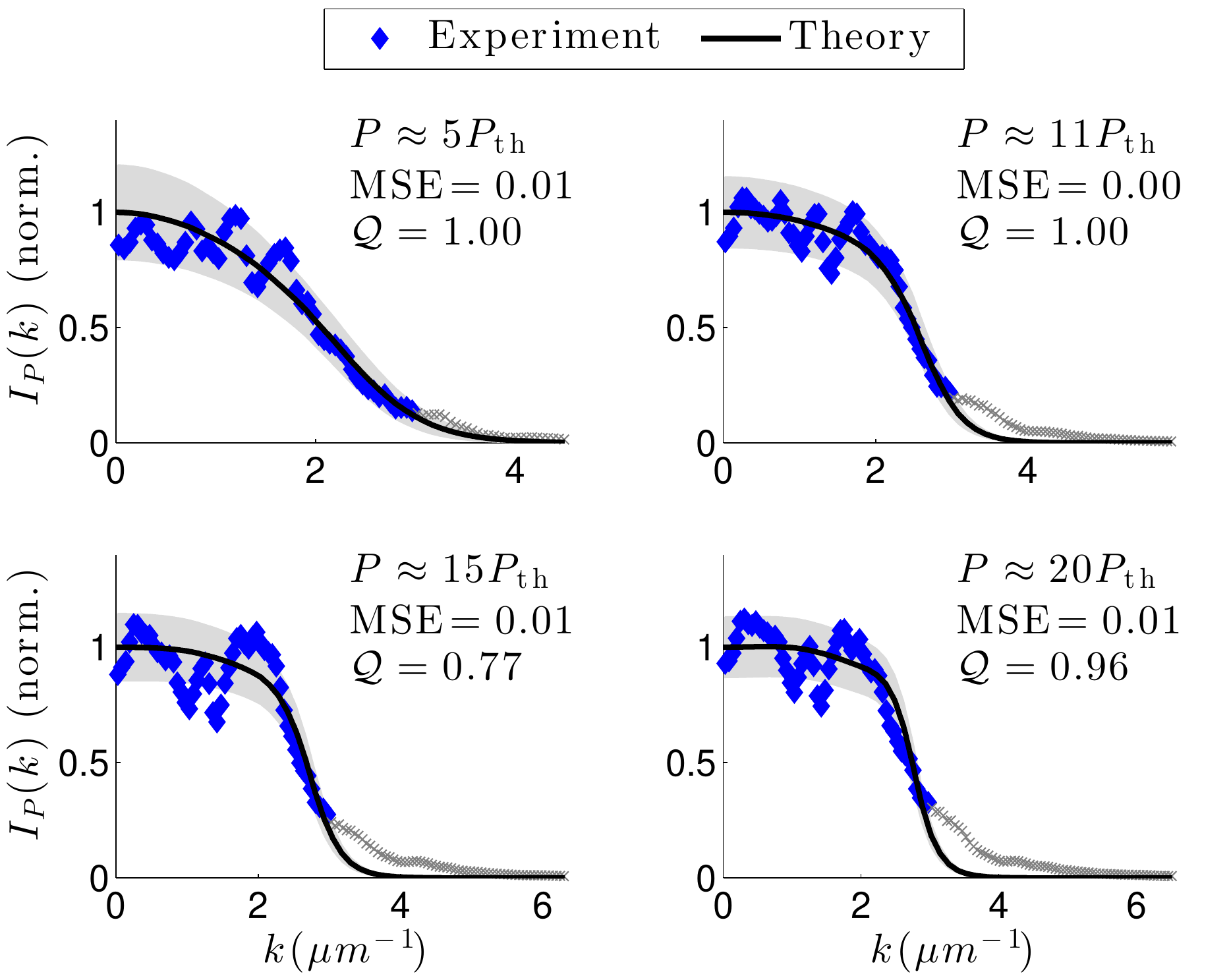}
	\caption{(color online)
Experimental $I_P(k)$ distribution (blue diamonds) compared to numerical simulations of scenario~II for increasing excitation power~$P$. The black solid lines depict the expectation value whereas the gray band indicates the standard deviation. For high momenta we have excluded systematically biased data (gray crosses). For each plot the mean squared error (MSE) and the goodness-of-fit value~$\mathcal{Q}$ (see text) are computed.
For used parameters see Tab.~\ref{tab:parameters}.
}
	\label{fig:Fit}
\end{figure}

For a quantitative analysis we compare directly the experimental measurements with the numerically computed expectation value~$\mu_{\mt{P}}$ and variance~$\sigma^2_{\mt{P}}$ of the intensity distribution. To this end we symmetrize the experimental data $I_P(k) \to (I_P(k) + I_P(-k))/2$ with $k \geq 0$, and superimpose them with the results of the numerical simulations. Since the condensate density and healing length are hard to determine experimentally, we fix the scaling of $x$- and $y$-axis by a least-square fit. 
Fig.~\ref{fig:Fit} shows the result. We have excluded experimental data with wavevectors~$k \geq 3 \: \mu \mt{m}^{-1}$, because a systematic artifact is present for all $k = 3,\dots,4 \ \mu \mt{m}^{-1}$ and for all excitation powers. \footnote{The shoulder in the experimental data at $k = 3,\dots,4 \ \mu \mt{m}^{-1}$ is visible for all excitation powers and its shape and magnitude is independent on the power. Therefore we can conclude that this shoulder is caused by residual effect and does not belong to the condensate emission. Different options are possible, e.g. the emission of uncondensed polaritons occupying the LP branch or artifacts from the experimental setup, e.g. transmission fluctuations from the microscope objective.}
%
%
In order to quantify the agreement between theory and experiment we introduce the mean squared error (MSE) and the goodness-of-fit value~$\mathcal{Q}$ (see Appendix~\ref{sec:theory_numsim} for definitions). $\mathcal{Q}$~is a probability: if $\mathcal{Q} \sim 1$, the simulations describe the experimental data. On the other hand, if $\mathcal{Q} \ll 1$ the theoretical model does not reproduce the experiment.
The experimental data are well described by simulations (of scenario~II), cf. Fig.~\ref{fig:Fit}: the MSE is close to zero, and the goodness-of-fit value~$\mathcal{Q}$ remains comparable to one for all pump powers studied.
In contrast, trying to reproduce the experimental observations by simulations of a polariton BEC described by scenario~I (quasi-equilibrium) fails, cf. Fig.~\ref{fig:FitScenarioI} of the Appendix~\ref{sec:theory_numsim}. Thereto, we had chosen a non-equilibrium parameter~$\alpha = 0.5$ and slightly increased the disorder strength. Then, the $\mathcal{Q}$-value drops from $\mathcal{Q} \approx 10^{-8}$  at $P\approx 5~P_{\mt{th}}$ to $\mathcal{Q} \approx 10^{-21}$ at $P = 20~P_{\mt{th}}$.

\section{Summary and Conclusion}
\label{sec:conclusion}

In this work we have characterized a polariton condensate in a disordered environment. Our theoretical analysis shows that spatial fluctuations of the condensate phase, which are induced by the interplay of disorder and gain-loss of particles, do not depend on the mean condensate density. This leads to a reduced stabilization against disorder fluctuations with increasing density in contrast to an equilibrium condensate.
%
%
To verify our prediction we have analyzed experimentally the photoluminescence emission of a ZnO based microcavity. Indeed, we find a lack of stabilization with increasing density in terms of pronounced intensity fluctuations within the $k$-space emission pattern even at high excitation power. This experimental finding can be reproduced by numerical simulations. From this we conclude that the polariton condensate in the microcavity is exposed to significant structural disorder, and that the persistence of disorder effects even at high excitation power, well above the condensation threshold, relies on the intrinsic non-equilibrium nature of polaritons.
We note that these findings may also explain the observation of similar phenomena for polariton condensates in microcavities based on other materials, e.g. CdTe or GaN.~\cite{Christopoulos.2007,Baas.2008,Krizhanovskii.2009}

\section{Acknowledgement}
We acknowledge experimental support and fruitful discussions from the group of N. Grandjean at EPFL. AJ is supported by the Leipzig School of Natural Sciences BuildMoNa. This work was supported by Deutsche Forschungsgemeinschaft through project GR 1011/20-2, by Deutscher Akademischer Austauschdienst within the project PPP Spain (ID 57050448) and by Spanish, MINECO Project Nos. MAT2011-22997 and MAT2014-53119-C2-1-R.\\
MT and AJ contributed equally to this work. MT performed the experimental part and AJ carried out the theoretical analysis.

{
\appendix


\section{Experimental setup}
\label{sec:sample_setup}



In order to investigate the optical properties of the polariton condensate, we applied two different photo-luminescence configurations, which have in common a non-resonant and pulsed excitation as well as a detection of the far-field emission. The setup to investigate the disorder effects on the polariton distribution and their dynamics as a function of the excitation power is described in Ref.~\onlinecite{Franke.2012}. Here, the excitation was carried out by a pulsed Nd:YAG laser with pulse duration of 500~ps, whose Gaussian excitation spot covers a sample area of about 10~$\mu$m$^2$. 

For the coherence measurements, the sample was excited via a frequency-tripled Ti:sapphire laser at 266~nm (repetition rate: 76~MHz, pulse length $\approx$ 2~ps). The PL signal of the Fourier plane was sent to a Michelson interferometer in the mirror-retroreflector configuration. The retroreflector image is a centrosymmetric counterpart of the mirror arm image.  In the resulting interferogram we superimposed the signal with wavevector $\vec{k_{||}}$ with that of $-\vec{k_{||}}$. Interference maxima occur when the path difference between the individual beams,  $\Delta L$~=~$c \Delta t$, is an integer multiple of the PL emission wavelength, being~$\Delta t$ the delay between the beams and $c$ the speed of light. With the help of a streak camera the relative delay between the two arms was set to zero for $k_{||} = 0$. 

Real space measurements with a sufficient spatial resolution could not be performed due a to a spherical aberration induced by the cryostat window. For the conditions used in our experiments, namely the UV spectral range, a window thickness of 1.5~mm and the large range of collected emission angles of $\pm 23^\circ$, the resulting spatial distortion of the image is larger than structural fluctuations that we would like to resolve. Consequently, the distortion of the measured real space image prevents a precise investigation of the spatial distribution of the luminescence as well as spatially resolved correlation measurements. We note that far-field images are not affected by the cryostat window, which causes a parallel beam shift but does not change the angle of the transmitted rays.

\section{Origin of disorder potential}
\label{sec:experiment_disorderorigin}

\begin{figure}
	\centering
	\includegraphics[width=0.3\textwidth]{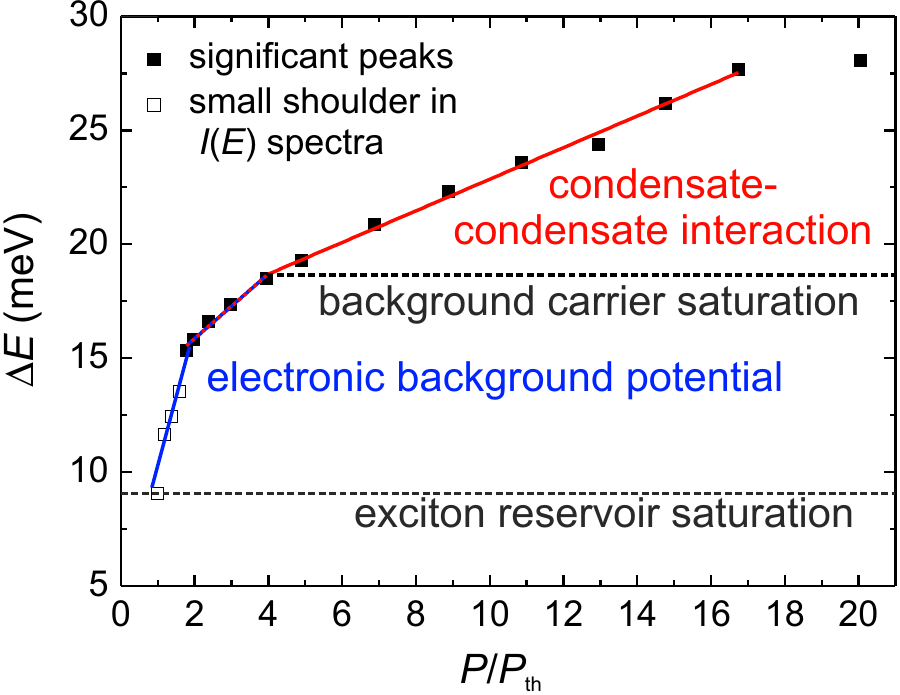}
	\caption{(color online)
Polariton blueshift $\Delta E$ as a function of the excitation power normalized to the condensation threshold for the observed ZnO MC. The red line corresponds to the polariton-polariton interaction whereas the blue line represents the blueshift due to an additional electronic background potential, which starts to saturated at about 2 $P_{\mt{th}}$ and is presumably totally saturated for $P > 4 \Pth$ .} 
	\label{fig:Blueshift}
\end{figure}

Due to the dual light-matter nature of the polaritons, the effective disorder potential can be of photonic as well as electronic origin. 




Photonic disorder can be caused by surface and interface roughness as well as thickness fluctuations within the MC structure. This leads to a spatial fluctuating cavity length and therefore to a variation of the cavity photon energy. Due to results of other ZnO-based MCs, a minimum potential strength of $V_C \geq$ 2 meV can be expected.~\cite{Li.2013,Orosz.2012} The corresponding correlation length~$\xi_V$  is of the order of the photonic wavelength, of about 370~nm.

In the literature, usually electronic disorder is neglected.~\cite{Trichet.2013,Baas.2008,Savona.2007,Manni.2011,Nardin.2009} In contrast to this, we assume a strong influence of an electronic background potential caused by randomly distributed excitonic states which are accumulated within the bulk of grains~\cite{Orton.1980} or bound to impurities. This is supported by two facts: firstly, cross-sectional TEM analysis of a MC that is fabricated under the same conditions, provides a granular structure of the investigated ZnO MC with grain sizes ranging from 20~nm up to 120~nm.~\cite{Franke.2012} Secondly, the slope of the polariton blueshift $\Delta E(P)$ is by a factor of about 6.3 larger for $P$~<~2~$P_{\textnormal{th}}$ than above and even by a factor of 12.6 larger compared to the blueshift for  $P > 4~\Pth$ (cf. Fig.~\ref{fig:Blueshift}). This can be explained by assuming an additional electronic background potential $\Delta E_b$, which may include localized states within a disorder potential or bound to impurities, as shown in Ref.~\onlinecite{Franke.2012,Franke.2012b}. Since the concentration of these electronic defects is finite, their contribution to the condensate blueshift saturates for a certain excitation power or rather condensate density. Thus, the further blueshift for $P$~>~4~$P_{\textnormal{th}}$ is restricted to condensate-condensate interaction.

We assume that the condensate blueshift for small excitation power $\Delta E$($P$~<~2~$P_{\textnormal{th}})$ is primarily caused by its interaction with aluminum donor bound excitons ($D^0,X$) and that $\Delta E$ scales linearly with its concentration. As mentioned in Sec.~\ref{sec:experimental_results}, we suppose a depletion of bound excitons at grain boundaries and thus an accumulation of them within the grain bulk.~\cite{Orton.1980} According to the model described in Ref.~\onlinecite{Orton.1980} the grain boundaries act like two back-to-back Schottky barriers and the carrier flow between grains is driven by thermionic emission over the Schottky barrier. In general, the average height and width of these barriers can be determined from the temperature-dependent evolution of the hall mobility. Unfortunately, this was not possible for our MC due to low current values, below the resolution limit of 1~nA, for temperatures below 200~K, caused by the small cavity thickness of about 100~nm as well as due to strong inhomogeneities of the current density, which may be caused by the cavity thickness gradient.

Assuming the mechanism of carrier depletion at grain boundaries to be the dominant one for the effective electronic disorder potential, its correlation length $\xi_V$ is similar to the grain size with values between 20~nm and 120~nm. This is about two orders of magnitude below the condensate size $L_c$, limited by the size of the pump spot and thus even lower than the assumed correlation length for photonic disorder of about 370~nm. Consequently, a trapping of the entire condensate within a minimum of the disorder potential can be excluded. We rather suppose that the disorder potential causes condensate density fluctuations and thus phase fluctuations due to the interplay of disorder and the non-equilibrium nature of the polariton condensate.

\section{Details of the Model}
\label{sec:theory_detailsmodel}
%
A phenomenological description of the dynamics of the macroscopic polariton condensate wave function~$\Psi(\vec{x},t)$ is given by an extended Gross-Pitaevskii Equation (eGPE),~\cite{Wouters.2007,Keeling.2008}
\begin{align}
	\nonumber
	\im \hbar \pd{t} \Psi = &\left( -\frac{\hbar^2}{2 m} \vec{\nabla}^2 + V(\vec{x}) + U \left|\Psi\right|^2 \right)\Psi \\
	\label{eq:eGPEdim}
		&+ \im \left(R(\vec{x}) - \Gamma \left|\Psi\right|^2\right) \Psi \ .
\end{align}
The first part of the right hand side is the ordinary equilibrium GPE with~$m$ as effective polariton mass of the lower polariton branch, $V(\vec{x})$ as external potential, and $U$ as repulsive onsite interaction potential. The second part models phenomenologically the gain and loss of condensed polaritons. Here, $R(\vec{x})$ describes the linear part of gain and loss, and the non-linearity implements a density dependent gain saturation with $\Gamma$ as gain depletion parameter.
This provides a simplified description of the gain process from a reservoir, for example relaxation of high-momentum polaritons generated by incoherent excitation with an external laser beam, and the condensate decay due to its finite lifetime.
Since the non-condensed polaritons have a short lifetime as compared to the lifetime of the condensate, we can safely neglect diffusion processes of these and relate the spatial extension of the reservoir to the Gaussian excitation profile of the laser beam. Then,
\begin{align}
	\label{eq:reservoir}
	R(\vec{x}) = \hbar \gamma_{\mt{c}} \left(\frac{P}{P_{\mt{th}}} e^{- \vec{x}^{\:2}/\xi_P^2} -1 \right) \ ,
\end{align}
with decay rate $\gamma_{\mt{c}} = 1/\tau$, where $\tau$ is the condensate lifetime, and waist size~$\xi_P$ of the laser beam. The parameter ~$P/P_{\mt{th}}$ is the excitation power normalized by its value at the threshold at which condensation is observed first.
The disorder landscape is incorporated by the external potential $V(\vec{x})$. We use a $\delta$-correlated Gaussian distributed quenched disorder with vanishing mean value and variance,
\begin{align}
	\label{eq:DisCorr}
	\expec{V(\vec{x})} = 0\ , \quad \expec{V(\vec{x})V(\vec{y})} = V_0^2 \xi_V^2 \delta(\vec{x}-\vec{y}) \ ,
\end{align}
respectively. The average disorder strength is given by $V_0$ and its characteristic length is denoted by $\xi_V$.

In the case of a spatially homogeneous excitation, i.e. $\xi_{\mt{P}} \to \infty$, our model~\eqref{eq:eGPEdim} was first suggested in Ref.~\onlinecite{Keeling.2008}.
As compared to Ref.~\onlinecite{Wouters.2007} we do not consider the dynamics of the reservoir polaritons explicitly. However, the latter can be eliminated~\cite{Carusotto.2013} for the typical case that the characteristic relaxation rate of the reservoir is much faster than the condensate decay rate~\cite{Wouters.2008,Carusotto.2013}. Then, an expansion to leading order in condensate density over reservoir density results in the eGPE~\eqref{eq:eGPEdim}.
We note that a different theoretical approach may be suitable in order to describe propagation of a polariton BEC in a disorder-free environment~\cite{Wertz.2012,Malpuech.2014}, which is not the aim of our work.

In the following we will discuss the model~\eqref{eq:eGPEdim}.
The mean condensate density $n_0\equiv \frac{1}{\Omega_{\mt{c}}} \int_{\Omega_{\mt{c}}} |\Psi(\vec{x})|^2$ is found by averaging the second term of the right hand side of \eqr{eq:eGPEdim} over the condensate area $\Omega_{\mt{c}} \approx \pi \xi_{\mt{P}}^2$, and then demanding a balance of gain and loss,
\begin{align}
	\label{eq:defn0}
	n_0 \approx \frac{\hbar \gamma_{\mt{c}}}{\Gamma} \left( \frac{P}{P_{\mt{th}}} -1 \right) \ .
\end{align}
Since the interaction term in \eqr{eq:eGPEdim} is proportional to the density, we find an energy blueshift~$n_0 U$. The healing length is extracted by comparing the kinetic energy term versus the interaction term in \eqr{eq:eGPEdim},
\begin{align}
  \label{eq:defhealinglength}
	\xi \equiv \sqrt{\frac{\hbar^2/2m}{n_0 U}} \ .
\end{align}
Let us understand its physical relevance: For example, we assume a region in which the condensate has to vanish, $\Psi = 0$, however remains unperturbed everywhere else. Then, the healing length is the distance over which the condensate density changes from zero to $n_0$.

A dimensionless eGPE~\eqref{eq:eGPEdim} takes the form
\begin{align}
	\label{eq:eGPEdimless}
	\im \partial_t \psi = (-\nabla^2 + \vartheta(\vec{x}) + |\psi|^2) \psi + i \alpha( g_\mt{R}(\vec{x}) - |\psi|^2)\psi \ ,
\end{align}
where density, length, energy and time are measured in units of $n_0$, $\xi$, $n_0 U$ and $\hbar/n_0U$, respectively. The 'non-equilibrium' parameter $\alpha$ and the dimensionless reservoir function $g_\mt{R}$ are defined in \eqr{eq:defkappaalpha} and \eqr{eq:defgP}, respectively. With  $\psi(\vec{x},t) \equiv \Psi(\vec{r},t)/\sqrt{n_0}$ we denote the dimensionless wave function and $\vartheta(\vec{x}) = V(\vec{x})/ n_0U$ is the disorder potential relative to the blueshift with
\begin{align}
  \label{eq:disorder_stat}
	\expec{\vartheta(\vec{x})} = 0 \ , \qquad \expec{\vartheta(\vec{x})\vartheta(\vec{y})} = \kappa^2 \delta(\vec{x}-\vec{y}) \ .
\end{align}
We have introduced two important dimensionless parameters, namely an effective disorder strength and a 'non-equilibrium' parameter
\begin{align}
  \label{eq:defkappaalpha}
  \kappa \equiv \frac{\xi_V \: V_0}{\xi\: n_0U} \ , \quad \text{and} \quad \alpha \equiv \frac{\Gamma}{U} \ .
\end{align}
The first parameter~$\kappa$ is also obtained by coarse graining the random disorder potential up to the healing length (assuming $\xi_V \ll \xi$). This process renormalizes the disorder strength by a factor~$1/\sqrt{(\xi/\xi_V)^2}$. Then, the value~$\xi_V V_0 / \xi$ is compared to the blueshift~$n_0 U$.
The second parameter~$\alpha$ implements the non-equilibrium nature of polaritons. In the limit $\alpha \to 0$ (keeping $n_0$ constant) \eqr{eq:eGPEdimless} reduces to the equilibrium GPE, whereas, for $\alpha \to \infty$ the condensate is totally dominated by gain and loss.
The rescaled  reservoir function yields
\begin{align}
	\label{eq:defgP}
	g_\mt{R}(\vec{x}) =  \frac{(P/P_{\mt{th}}) \: e^{-x^2/x_P^2}-1}{P/P_{\mt{th}}-1} \ ,
\end{align}
with $x_{\mt{P}} \equiv \xi_P/\xi$.
For a steady state solution (single-mode condensate) we make the ansatz
\begin{align}
  \label{eq:ssansatz}
	\psi(\vec{x},t) = \psi(\vec{x}) e^{-\im \omega t} = \sqrt{n(\vec{x})} e^{\im \phi(\vec{x}) - \im \omega t} \ ,
\end{align}
where $\hbar \omega$ is the condensate energy.

We emphasize that both blueshift and healing length  depend on the excitation power $P$  via $n_0$. Thus, $\kappa$ and $x_P$ depend on $P$, too. For our analysis it is useful to identify energy and length scale which are excitation power independent, namely the line width energy $\hbar \gamma_{\mt{c}}$ and the quantum correlation length $l_\gamma \equiv \sqrt{\hbar / 2 m \gamma_{\mt{c}}}$ (a non-equilibrium analogon of the thermal de Broglie wavelength)~\cite{Trichet.2013}, so that $\kappa$ and $x_P$ become functions of $\alpha, P/\Pth$ and sample parameters (see Table~\ref{tab:parameters}).
%
%

\section{Numerical Simulations and Comparison with the Experiment}
\label{sec:theory_numsim}
%
\emph{Numerical simulations --}
Computing the condensate wave function by solving the eGPE~\eqref{eq:eGPEdim} allows us to extract the real and $k$-space intensity distribution.
We define the $k$-space intensity distribution according to
\begin{align}
  \label{eq:defIk}
        I_P(\vec{k}) &\equiv \gamma_{\mt{c}} n_0 \: |\psi({\vec{k}})|^2\ ,
\end{align}
where the momentum space wave function is defined via a two-dimensional discrete Fourier transform $\psi(\vec{k}_j) = (1/N^2) \sum_{\vec{x}_i} \psi(\vec{x}_i) e^{-\im \vec{k}_j \vec{x}_i}$,  with $\vec{x}_i,\vec{k}_j$ being elements of a discrete lattice with $N$ lattice points in each spatial direction, such that $i,j = 1,\ldots,N^2$.
We choose an appropriate set of simulation parameters extracted from the experiment (see Table~\ref{tab:parameters}) and solve \eqr{eq:eGPEdimless} numerically. To this end we look for a steady state solution, see \eqr{eq:ssansatz}, by solving the time evolution of the discretized wave function $\psi(\vec{x}_i,t)$. The latter is defined on a real-space square lattice with spacing~$a=\xi$. We employ a variable order Adams-Bashforth-Moulton PECE algorithm~\cite{NR.2007} to obtain the time evolution. First, we compute the steady state solution of the disorder-free system, $\vartheta = 0$. Then, we choose independent Gaussian distributed variables of vanishing mean and variance~$\kappa^2$ for each lattice site and calculate the steady state solution of the disordered system. The time evolution of the disordered system is started from the disorder-free solution as initial condition. For each disorder realization the discretized two-dimensional wave function~$\psi(\vec{x}_i)$ of the steady state is extracted, and then Fourier transformed in order to compute the two-dimensional $k$-space intensity $I_P(\vec{k}_j)$. Finally, we average over all disorder realizations and compute the expectation value and variance,
\begin{align}
	\label{eq:meanIk}
	\mu_P(k) &\equiv \expec{I_P(\vec{k})}/\expec{I_P(0)} \ ,\\  
	\label{eq:varIk}
	\sigma^2_P(k)  &\equiv \expec{\left( I_P(\vec{k})-\expec{I_P(\vec{k})} \right)^2}/\expec{I_P(0)}^2\ ,
\end{align}
respectively. Above, the bracket $\expec{\ldots}$ denotes an average with respect to disorder, and we have normalized the mean and variance by the expectation value of the intensity at $k=0$. Since the excitation profile \eqr{eq:defgP} is radial symmetric, Eqs.~(\ref{eq:meanIk},~\ref{eq:varIk}) are radial symmetric, too, assuming a sufficiently large number of disorder realizations.

\begin{figure}[tb]
	\centering
	\includegraphics[width=1.00\hsize]{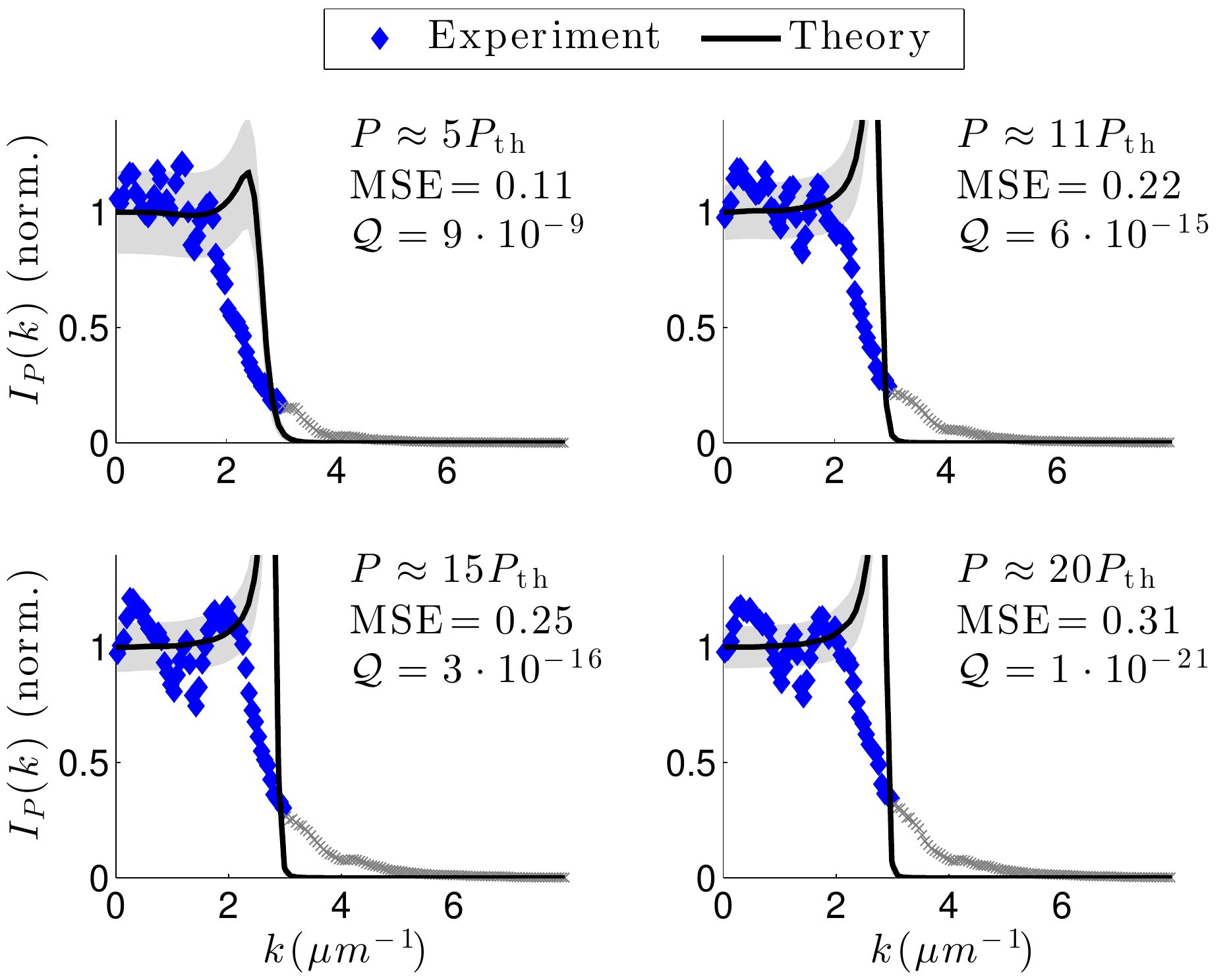}
	\caption{(color online)
Experimental $I_P(k)$ distribution (blue diamonds) compared to numerical simulations of scenario~I for increasing excitation power~$P$. The black solid lines depict the expectation value whereas the gray band indicates the standard deviation. For high momenta we have excluded systematically biased data (gray crosses). For each plot the mean squared error (MSE) and the goodness-of-fit value~$\mathcal{Q}$ are computed.
Used simulation parameters: $\alpha = 0.5$, $\xi_{\mt{V}} V_0 / l_{\mt{c}} \hbar \gamma_{\mt{c}} = 0.4$, $\xi_P/l_{\mt{c}}=3$. These correspond to $\Ln/\Lc \approx 1.5$, $\Lphi/\Lc \approx 13$, $\Ls/\Lc \approx 26$.
}
	\label{fig:FitScenarioI}
\end{figure}
\emph{Comparison with the experiment --}
The numerically obtained  mean and variance of the $k$-space intensity can be compared  with the experimental data denoted by $I_{\mt{ex}}(k_x)$ here, cf. Sec.~\ref{sec:experimental_results}. We note that these measurements represent a line-cut along an axis (e.g.~the $x$-axis) of the two-dimensional intensity distribution and are measured for one disorder configuration determined by the disorder of the sample. We perform a spatial averaging step by symmetrizing the experimentally obtained intensity:  $I_{\mt{ex}}(k_x) \to (I_{\mt{ex}}(k_x) + I_{\mt{ex}}(-k_x))/2$ and $k_x \geq 0$.
In order to quantify the agreement between experiment and theory we introduce the chi-square value~\cite{NR.2007}
\begin{align}
	\label{eq:DefChiSqr}
	\chi^2_P = \sum \limits_{k_j \geq 0} \left( \frac{I_{\mt{ex}}(k_j)/a - \mu_{P}(b k_j)}{\sigma_{P}(b k_j)} \right)^2 \ .
\end{align}
Since the condensate density~$n_0$ and the healing length~$\xi$ are hard to extract experimentally, we use two scaling parameters ($a,b$) instead. Both are determined by a least-squares fitting procedure.~\cite{NR.2007}


In order to estimate the goodness-of-fit~\cite{NR.2007} we extract the complement of the $\chi^2$-probability distribution function $F_{\chi^2}$, denoted by $\mathcal{Q} = 1-F_{\chi^2}(\chi^2_P)$, which is the probability that the simulations agree with the experimental data. If $\mathcal{Q} \ll 1$, the apparent discrepancies of model and data are unlikely to be random fluctuations, and we conclude that the model is not specified correctly, or that the fluctuation strength~$\sigma_P$ is under-estimated. On the other hand, if $\mathcal{Q} \sim 1$, we conclude that the model describes the data correctly.
Finally, we define the mean squared error: $\mt{MSE} \equiv (1/N^2) \: \sum_{k_j} (I_{\mt{ex}}(k_j)/a - \mu_P(k_j))^2$, which is a measure of how well the data match the simulated intensity distribution.
The comparison of the experimental data and the numeric simulations of scenario~I is shown in Fig.~\ref{fig:FitScenarioI}, and the comparison with simulations of scenario~II was shown and discussed in Sec.~\ref{sec:fit_of_exp_data}, Fig.~\ref{fig:Fit}.

} 

%

\input{source_v2.bbl}
%

%
%
\clearpage
\setcounter{equation}{0}
\setcounter{figure}{0}
\setcounter{section}{0}
\renewcommand{\theequation}{S.\arabic{equation}}
\renewcommand{\thefigure}{S.\arabic{figure}}
\renewcommand{\thesection}{SM \arabic{section}}

\onecolumngrid
\section*{Supplemental Material}
\twocolumngrid
\section{Experimental Determination of Condensation Threshold Density}
\label{sec:SM_Pth}

For the comparison between the theoretical model and the experiment as discussed in Sec.~\ref{sec:fit_of_exp_data} of the main text the determination of the condensation threshold $\Pth$ is crucial, since experimental as well as numerically simulated spectra shall be compared for similar ratios of $P / \Pth$. 

\begin{figure}[hb]
\centering
\includegraphics[width=0.7\linewidth]{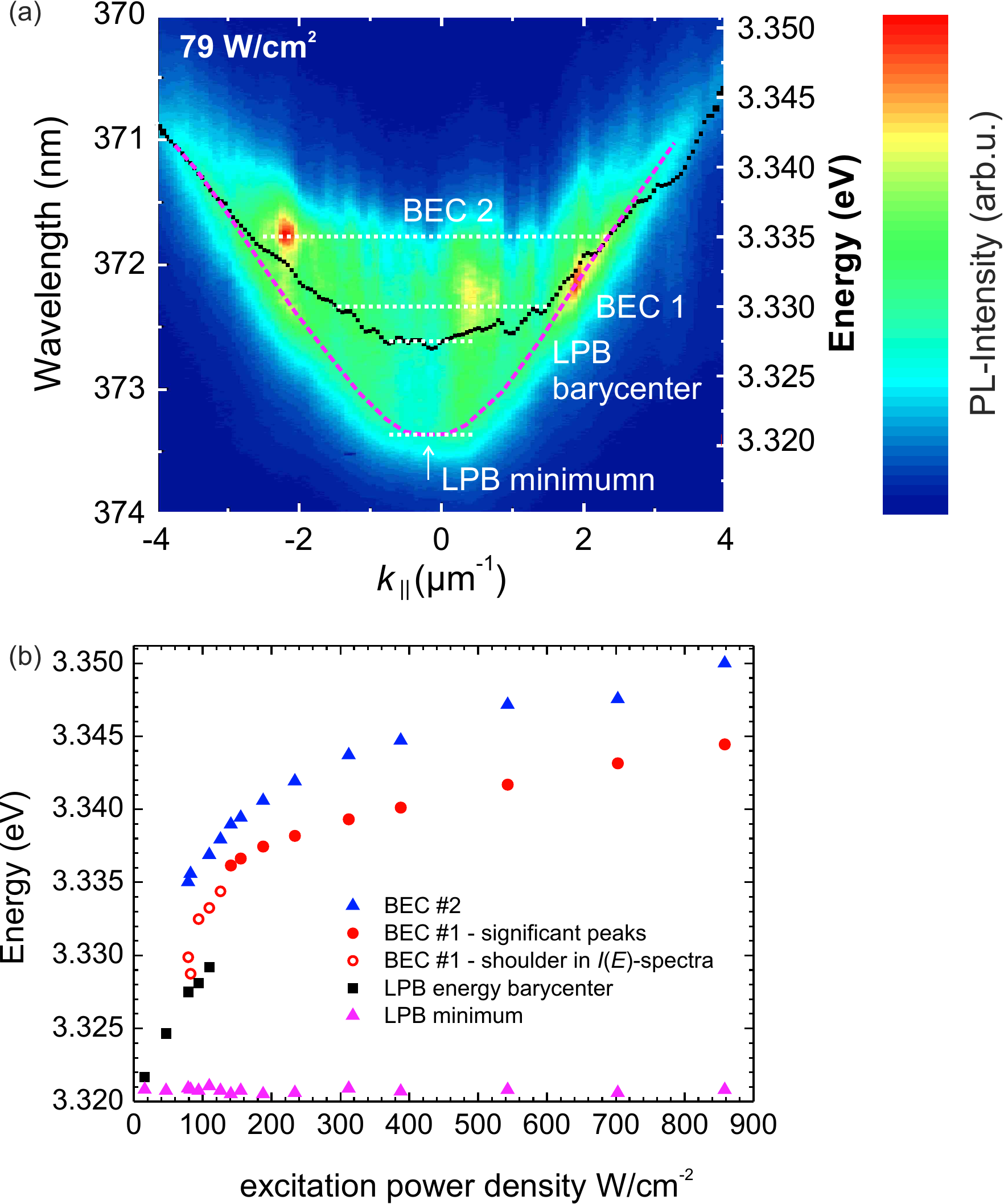}
\caption{(a)Far-field emission pattern $I(E,k)$ for $P =  79~\mt{W}/\mt{cm}^{2}$. The black dotted line represents the energy barycenter for the LPB emission $E^\mt{BC}_\mt{LPB}(k)$ for each $k$ value. The white dashed lines highlight all relevant spectral positions which are plotted for increasing excitation power density in (b). Here, for the LPB the minimum of the parabolic dispersion as well as the energy barycenter for $k \approx 0$ was analyzed. The empty red circles represent BEC emission that appears as shoulder within the PL intensity spectra leading to larger uncertainties for the determination of the peak energy. (cf. Fig~S.\ref{fig:SM:peakArea}(a))}
\label{fig:SM_Intro_spectr_contr}
\end{figure}
In general, we identify three spectral contributions within our far-field PL emission pattern, as shown in Fig.~S\ref{fig:SM_Intro_spectr_contr}, namely the lower polariton branch (LPB) emission, and two Bose-Einstein condensate (BEC) emission channels. Whereas the minimum of the LPB dispersion is almost excitation power independent ($E_\mt{LPB}^\mt{min} = 3.3207$~eV), the LPB emission gets significantly broadened towards higher energies for increasing excitation power. This is due to the fact that polaritons are located at different regions and are subject to different blueshifts due to the spatially inhomogeneous, e.g. Gaussian pump spot as well as the pronounced disorder potential. To consider this effect quantitatively, we calculated the energy barycenter of the LPB emission $E^\mt{BC}_\mt{LPB}(k) = (\int_{E} I(E) \cdot E \; dE) / (\int_{E} I(E) dE)$  for $k \approx 0$. To investigate the excitation power dependent evolution of the BEC emission, we performed a lineshape analysis of the PL spectra which are integrated over all observed $k$ values by assuming a Lorentzian lineshape. Both BEC emission channels show a large energy shift of about $E_\mt{BEC 1} = E_\mt{BEC 2} = 16\pm2$~meV with respect to $E_\mt{LPB}^\mt{min}$ for increasing excitation power density up to $P =  155~\mt{W}/\mt{cm}^{2}$. We assume that the initial large energy shift of $E^\mt{BC}_\mt{LPB}$ as well as of $E_\mt{BEC 1,2}$ is mainly caused by an electronic background potential that is discussed in detail in Appendix~\ref{sec:experiment_disorderorigin} in the main text. For $P >  155~\mt{W}/\mt{cm}^{2}$ the slope of $E_\mt{BEC 1,2}$ is reduced, following the expectations for common BEC. In this regime, the electronic background potential starts to saturate and the condensate-condensate interaction becomes dominant. 

Unfortunately, the experimental determination of the condensation threshold is accompanied by large uncertainties due to the interaction between both condensate modes as well as the superposition of intense LPB emission for a large range of excitation powers. Therefore, we discuss here the impact of the disorder on the determination of the threshold power. By doing so, we analyze at first the evolution of the total PL intensity with increasing excitation power, as usually done for a disorder free condensate. As will be discussed below, this method gives only an upper limit  and therefore we apply two further methods:  firstly, we examine the excitation power dependent evolution of the PL intensity for each spectral contribution separately. Secondly, we study the PL spectra $I_k(E)$ for each $k$ value and excitation power separately and deduce the FWHM for the individual emission channels as a function of $k$ and excitation power density.    
\\
\paragraph{Evolution of the total Photoluminescence (PL) Intensity}
\label{par:totalInt}

\begin{figure}[htb]
	\centering
\includegraphics[width=0.7\linewidth]{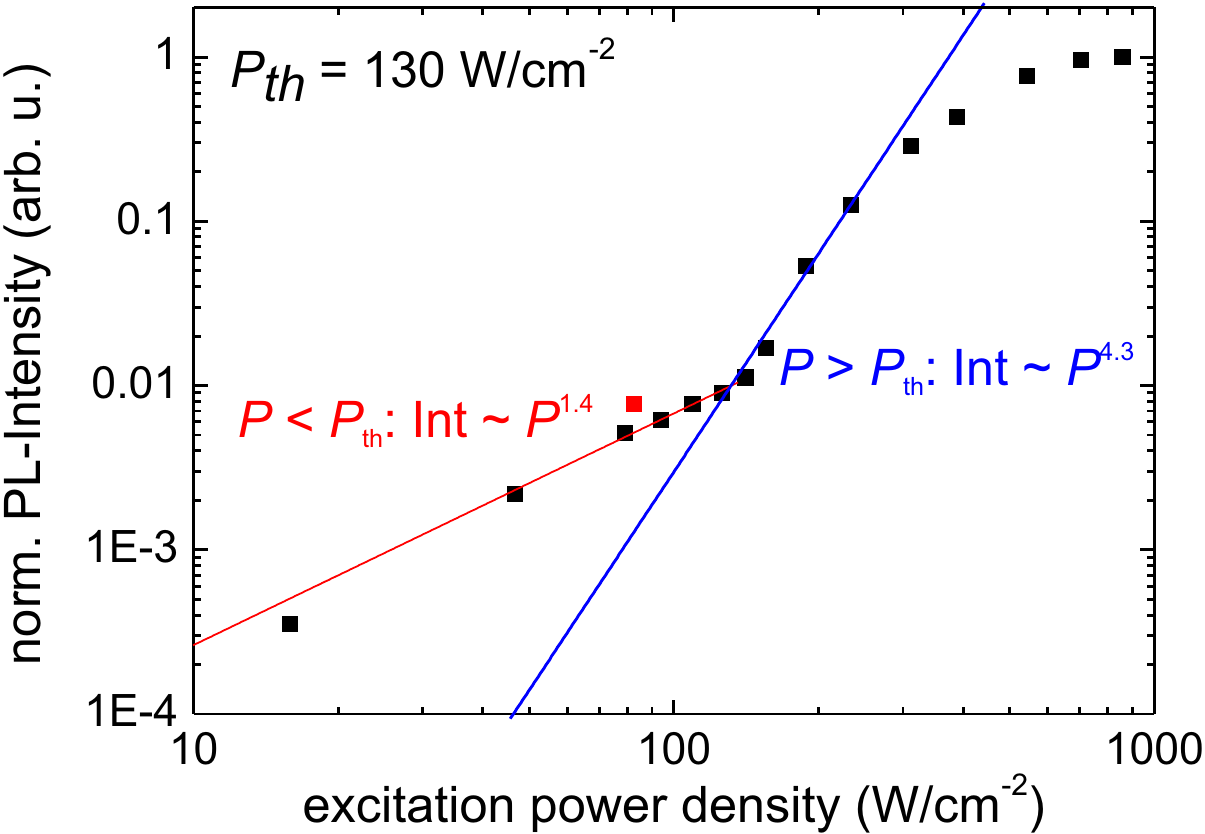}
	\caption{(color online)
		Excitation power dependence of the total PL intensity. For small excitation power densities $P \leq 235~\mt{W}/\mt{cm}^{2}$ the behavior can be described by two power functions with different exponents below (red solid line) and above (blue solid line) the condensation threshold. The threshold can be deduced from the intersection between both functions: $\Pth = 130\mt{W}/\mt{cm}^{-2}$.}
	\label{fig:SM:totalInt}
\end{figure}

As a first guess, we analyzed the excitation power dependence of the total PL intensity, integrated over all observed $k$ values and energies, as shown in Fig.~S\ref{fig:SM:totalInt}. The slope of the PL intensity increases abruptly for $P \geq 130~\mt{W}/\mt{cm}^{2}$. By assuming a power law behavior,\cite{kavokin.2007} the exponent increases from 1.4 for $P < 130~\mt{W}/\mt{cm}^{2}$ to 4.3 for $P > 130~\mt{W}/\mt{cm}^{2}$.  Note that the estimated value of $\Pth = 130~\mt{W}/\mt{cm}^{2}$ is only an upper limit for the condensation threshold. This can be explained by the inhomogeneous shape of the (e.g. Gaussian) excitation spot profile. For excitation powers $P \gtrsim \Pth$ the critical density for polariton condensation is achieved within a small area only. In contrast to this, emission from uncondensed polaritons occurs for a much larger area, which superimposes the BEC emission. Thus, the BEC emission becomes dominant leading to the observed kink in the evolution of the PL intensity with increasing powers only for powers significantly larger than the real condensation threshold.
\\
\paragraph{Evolution of PL Emission for each BEC State}
\label{subsec:BEC_peakArea}

\begin{figure}[htb]
	\centering
	\includegraphics[width=0.9\linewidth]{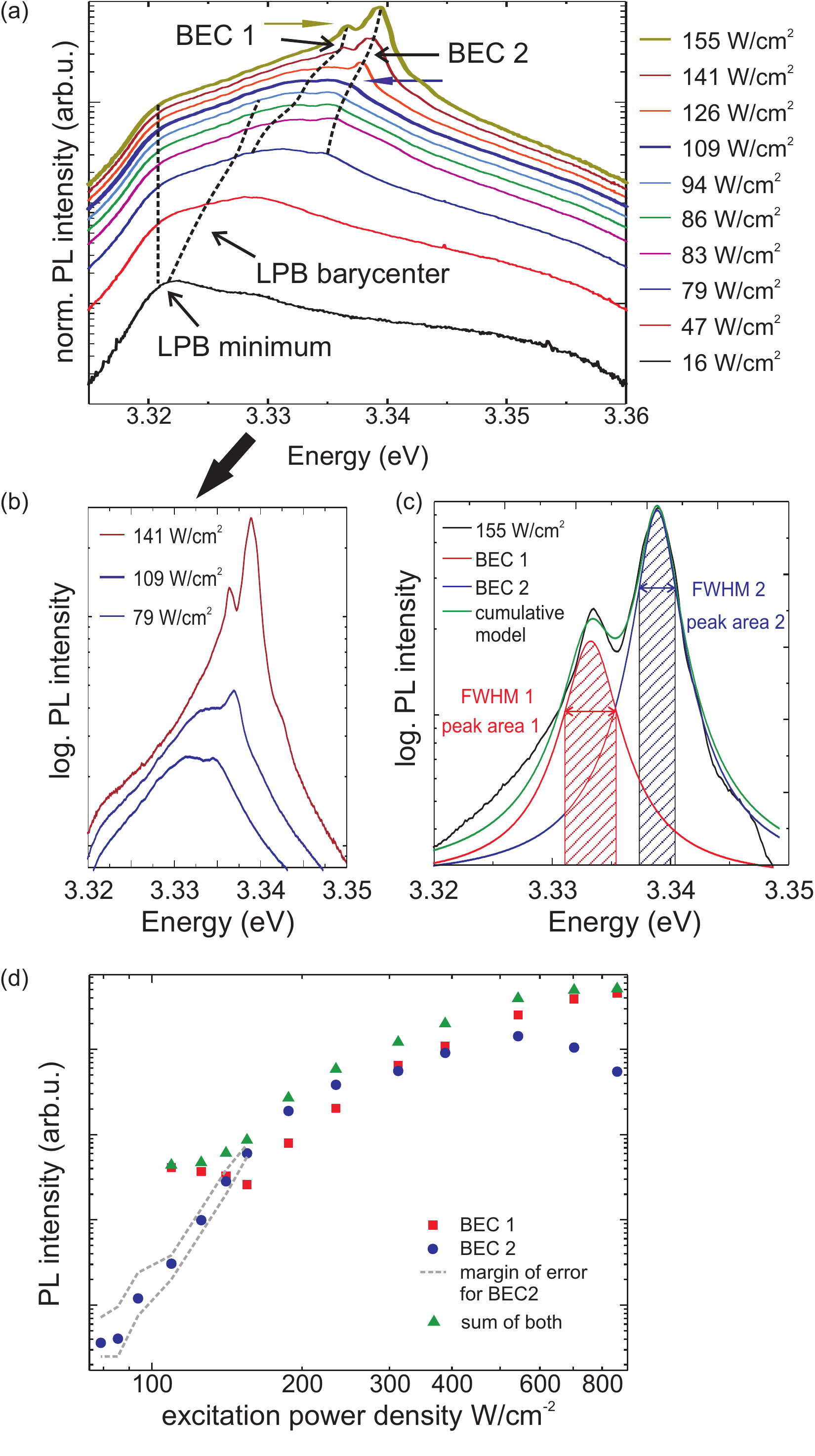}
	\caption{(color online)		
		(a) Normalized	PL spectra $I(E)$ integrated over all $k$ values for different excitation powers. The blue (dark yellow) arrow points at the PL spectra for which firstly a pronounced peak relating to the emission of BEC~2 (BEC~1) is observed. (b) Selected spectra that are significant for the determination of $\Pth$. (c) Exemplary PL spectrum for $P =  155~\mt{W}/\mt{cm}^{2}$ showing the FWHM of both BEC emission channels as well as the peak area. (d) Excitation power dependence of the PL intensity, integrated over the FWHM of the corresponding peak.}
	\label{fig:SM:peakArea}
\end{figure}

To analyze the impact of the superposition of the LPB and BEC emission on the determination of $\Pth$, we analyzed the excitation power dependent evolution of the PL emission for both contributions separately. For this purpose, we investigated the intensity spectra $I(E)$, integrated over all observed $k$ values, for different excitation power densities as shown in Fig.~S\ref{fig:SM:peakArea}(a). For the lowest density ($ P = 16~\mt{W}/\mt{cm}^{2}$) only the LPB emission can be observed, with maximum intensity at the minimum of the LPB dispersion of $E_{\mt{LPB}} = 3.3207~\mt{eV}$. For increasing excitation power, the high energy edge of the LPB emission dominates due to the previously mentioned significant broadening towards higher energies. Already for $P = 79~\mt{W}/\mt{cm}^{2}$ an additional emission channel appears at $E_\mt{BEC 2} = 3.335~\mt{eV}$. This peak becomes pronounced and shows a strong narrowing for $P \geq 109~\mt{W}/\mt{cm}^{2}$. For further increasing excitation power a second pronounced BEC emission channel appears within the $I(k)$ spectra at smaller energies indicating the multimode BEC behavior, as discussed in Sec.~\ref{sec:experimental_results} in the main text. Note that this emission channel is already observable for  $P = 79~\mt{W}/\mt{cm}^{2}$ in the energy resolved $k$-space images as shown in Fig.~S\ref{fig:SM_Intro_spectr_contr}(a). However, it appears only as a small shoulder within the $k$-integrated intensity spectra for this excitation density range (cf.~Fig.~S\ref{fig:SM:peakArea}(a,b)). The energy position of the spectral contributions considered here (cf. Fig.~S\ref{fig:SM_Intro_spectr_contr}(b)) is indicated by the dashed lines in Fig.~S\ref{fig:SM:peakArea}(a).

An exemplary lineshape analysis of the BEC~1 and BEC~2 contributions to the PL spectra is shown in Fig.~S\ref{fig:SM:peakArea}(c) for an excitation power of $P = 155~\mt{W}/\mt{cm}^{2}$. The dependence on excitation power density of the PL peak area, integrated over the FWHM of the corresponding Lorentzian peaks, for both BEC emission channels separately as well as for the sum of both is compiled in Fig.~S\ref{fig:SM:peakArea}(d) in a double-logarithmic scale.
The PL intensity of BEC~2 starts to saturate for $P \approx 300~\mt{W}/\mt{cm}^{2}$ and even decreases for $P >  550~\mt{W}/\mt{cm}^{2}$, whereas the PL intensity of BEC~1 further increases for the excitation power density range shown here. This indicates an effective relaxation of polaritons from the high-energy BEC state~2 into the low-energy BEC state~1. Thus, both BEC emission channels are not independent but represent a system of coupled condensate states for which we estimate a single condensation threshold density. Similar to the previous method, we expect a kink in the evolution of the PL intensity at $\Pth$, however, only a discontinuity is barely visible at $P \approx 84~\mt{W}/\mt{cm}^{2}$ for the data set presented here (cf. Fig.~S\ref{fig:SM:peakArea}(d)). We note that the PL spectra for $P <  109~\mt{W}/\mt{cm}^{2}$ do not show a clear peak for the energy range of the expected BEC emission due to the spectral overlap with the intense and spectrally broad LPB emission. Thus, the extracted peak area is subjected to large uncertainties in the mentioned range of excitation densities, which is highlighted by the gray dashed lines in Fig.~S\ref{fig:SM:peakArea}(d). Due to this fact, the observed discontinuity in the evolution of the PL intensity is not fully reliable and only a range of possible values can be determined for $\Pth$. On the one hand, a peak at 3.335~eV is slightly visible for an excitation density of $P =  79~\mt{W}/\mt{cm}^{2}$, which leads to the observed discontinuity in the PL intensity evolution  at $P \approx 84~\mt{W}/\mt{cm}^{2}$ and may indicate the onset of BEC emission. However, this peak may also be caused by effective polariton scattering into an already blueshifted state in the uncondensed regime. On the other hand, the peak gets pronounced and spectrally narrowed for  $P =  109~\mt{W}/\mt{cm}^{2}$. This is a clear signature for polariton condensation. Conclusively, the range of values for $\Pth$ can be restricted to $\Pth = (84 - 109)~\mt{W}/\mt{cm}^{2}$ by using this method. A way to additionally reduce the impact of spectral  overlapping between LPB and BEC emission and further specify $\Pth $ is discussed in the following section.             
\\

\paragraph{$k$-dependent Evolution of FWHM}
\label{par:FWHM_vs_P}  

\begin{figure}
	\centering
	\includegraphics[width=0.7\linewidth]{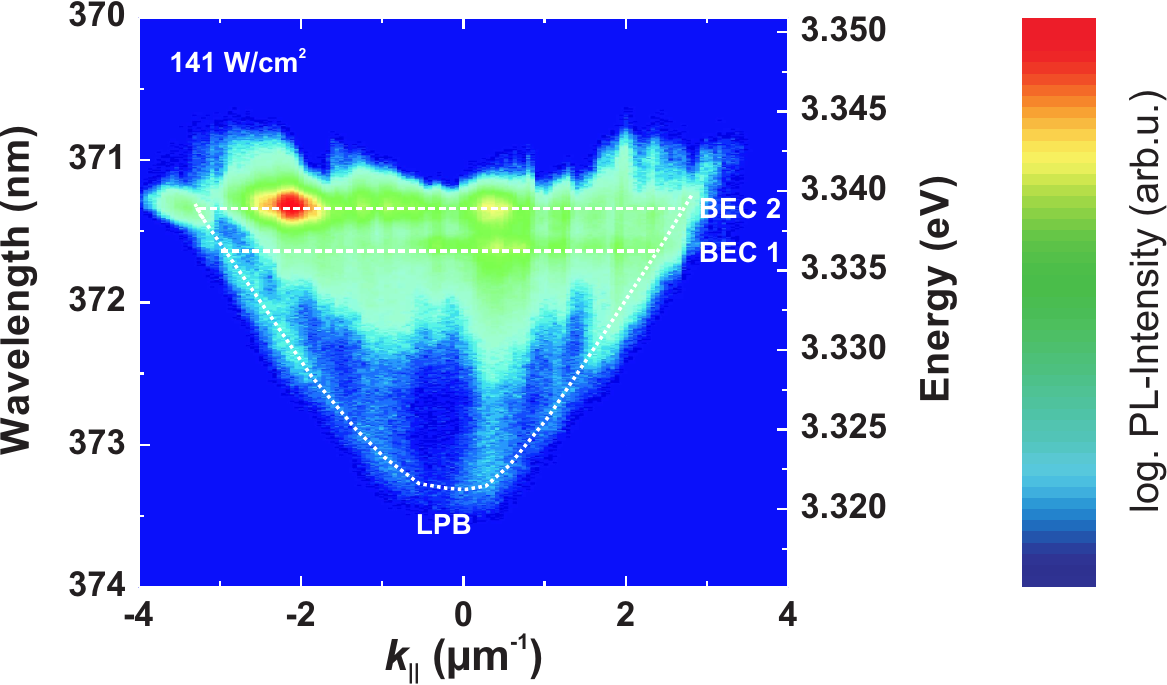}
	\caption{(color online)
	Far-field emission pattern $I(E,k)$ for an exemplary excitation power of $P =  141~\mt{W}/\mt{cm}^{2}$. The parabolic shaped LPB emission as well as the dispersionless BEC emission channels are highlighted by white dashed lines. }  
\label{fig:SM:PL_Image}
\end{figure}
   
For a more sophisticated investigation we analyzed the PL intensity spectra  $I_k(E)$ for each $k$ value and excitation power separately. Thereby, we deduced the FWHM for both BEC emission channels as well as for the LPB emission. Fig.~S\ref{fig:SM:PL_Image} shows exemplarily a far-field PL emission pattern for $P =  141~\mt{W}/\mt{cm}^{2}$ with logarithmic intensity scale. Here, all three emission channels, marked by white dotted lines, are energetically well separated for a large range of $k$ values.
\begin{figure}
	\centering
	\includegraphics[width=\linewidth]{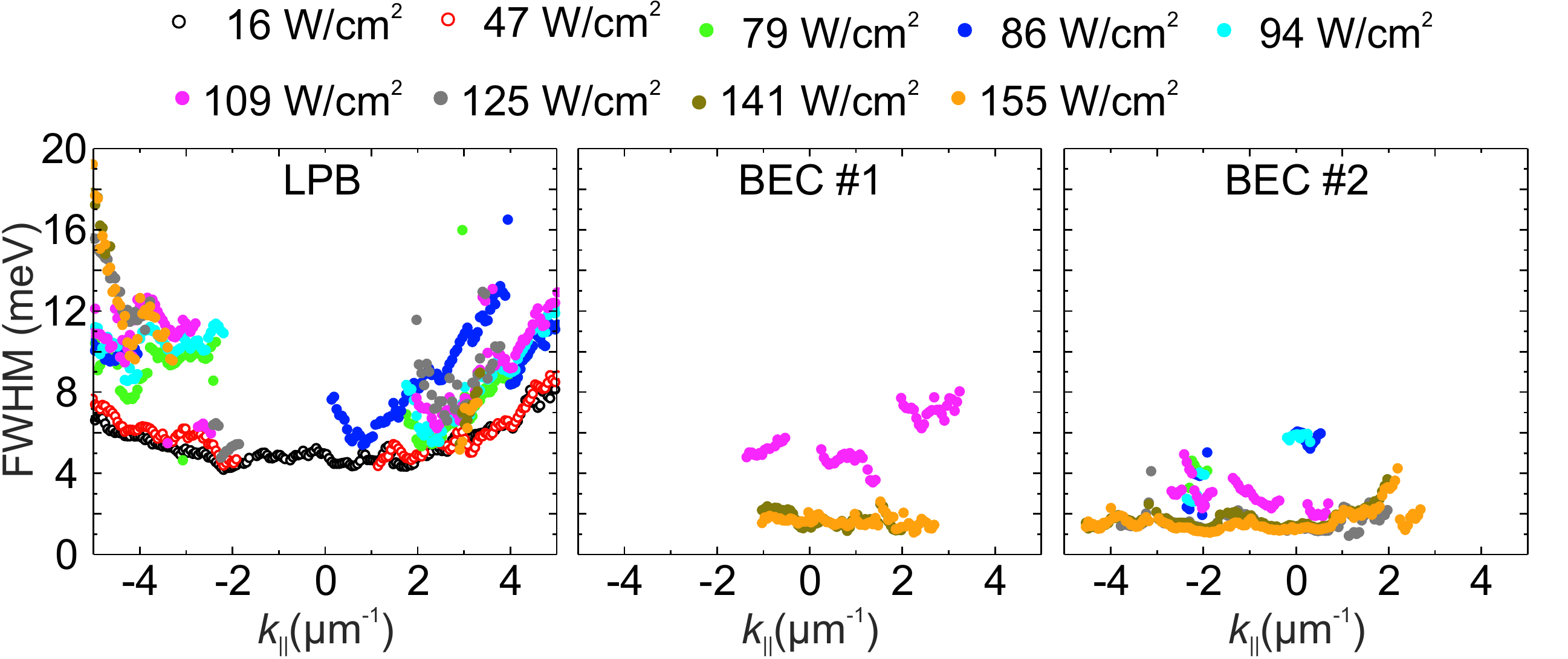}
	\caption{(color online)
		$k$-dependent FWHM of LPB and BEC emission for a large range of excitation power densities. Empty circles represent excitation power densities $P < \Pth$, whereas excitation power densities $P \geq \Pth$  are highlighted by full ones.}  
	\label{fig:SM:FWHM}
\end{figure}
The $k$-dependent evolution of the FWHM with increasing excitation power density is shown in Fig.~S\ref{fig:SM:FWHM} for each emission channel separately. Note, that the missing data points correspond to PL spectra which show a strong spectral overlap of the emission channels and thus prevent a proper lineshape analysis. The broadening of the LPB emission increases with increasing absolute $k$ values due to an increasing excitonic fraction. For $P \geqslant  79~\mt{W}/\mt{cm}^{2}$ the FWHM of the BEC peak 2 is lower than the minimum FWHM of the LPB emission indicating the onset of polariton condensation. For the BEC peak 1 this situation is present for $P \geqslant  109~\mt{W}/\mt{cm}^{2}$. For both BEC emission channels, the narrowing saturates for $P =  141~\mt{W}/\mt{cm}^{2}$, indicating a maximum temporal coherence. Following the arguments about the interaction between both investigated BEC emission channels as discussed in the previous section, this coupled BEC system is characterized by a single threshold power of $\Pth =  79~\mt{W}/\mt{cm}^{2}$. \\

\paragraph{Summary}

The results for the determination of $\Pth$ are summarized for all three methods in Table~\ref{tab:Pth}. Using the typically used method by analyzing the total PL intensity as a function of excitation power, an upper value of $\Pth <  130~\mt{W}/\mt{cm}^{2}$ was estimated. By studying the PL intensity for each spectral contribution separately  it was possible to reduce the impact of the disorder on the determined threshold value and further restrict the range of $\Pth$ to  $\Pth = (84- 109)~\mt{W}/\mt{cm}^{2}$. The best minimization of the disorder influence on the determination of the threshold power for condensation was achieved by investigating the PL spectra for each $k$ value separately and deducing the FWHM for each spectral contribution. This method also differs from the other ones regarding its physical principle. The analysis of the PL intensity evolution for the total emission as well as for the individual BEC emission are based on an increasing rate of the parametric scattering process into the condensate state for $P \geq \Pth$ due to its bosonic nature. Thereby, the coexistence of a small area of condensed polaritons for $P \gtrapprox \Pth$ and a large area of uncondensed ones can cause a rather soft transition of the PL intensity evolution leading to large uncertainties for the determination of $\Pth$. Investigating the FWHM of the BEC emission channels rather gives insight into another property. As the FWHM of the BEC emission channels are inversely proportional to the temporal coherence of the corresponding system of particles the spontaneous build-up of coherence is observed, which is a basic property of a polariton condensate. Nevertheless, the convolution of a certain BEC emission peak with other emission channels leads to uncertainties for the determination of $\Pth$ using this method, too. In summary, we estimate $\Pth =  79~\mt{W}/\mt{cm}^{2}$ as the threshold value for polariton condensation in our MC for the investigated parameter set of $T = 10~\mt{K}, \Delta = -30~\mt{meV}$.

\begin{table}[tb]
	\begin{tabular}{|p{2.5 cm}|p{1.2cm}|p{4.5 cm}| }
		\hline
		method & $\Pth$ in $\mt{W}/\mt{cm}^{2}$   & comments \\
		\hline
		total PL intensity & 130 & upper limit, BEC emission superimposed by intense LPB emission \\
		\hline
		PL intensity for each BEC emission channel  & 84 - 109 &
		superposition with LPB emission prevents reliable lineshape analysis for small excitation power densities \\
		\hline
		$k$-dependent FWHM  & 79 & provides best separation between LPB and BEC emission \\
		\hline
		%
	
		%
		%
	\end{tabular}
	\caption{Used methods for experimental determination of the condensation threshold density $\Pth$.}
	\label{tab:Pth}
\end{table}

\section{Experimental limitations for Determination of the Coherence Time}
\label{sec:SM_limits_cohTime}
In this section, we discuss two experimental artifacts that lead to limitations in the determination of the condensate's coherence time. 

\subsection{Spectrally dependent Phase Shift}
\label{subsec:Tcoh_phase_shift}
\begin{figure}[hbt]
	\centering
	\includegraphics[width=0.8\linewidth]{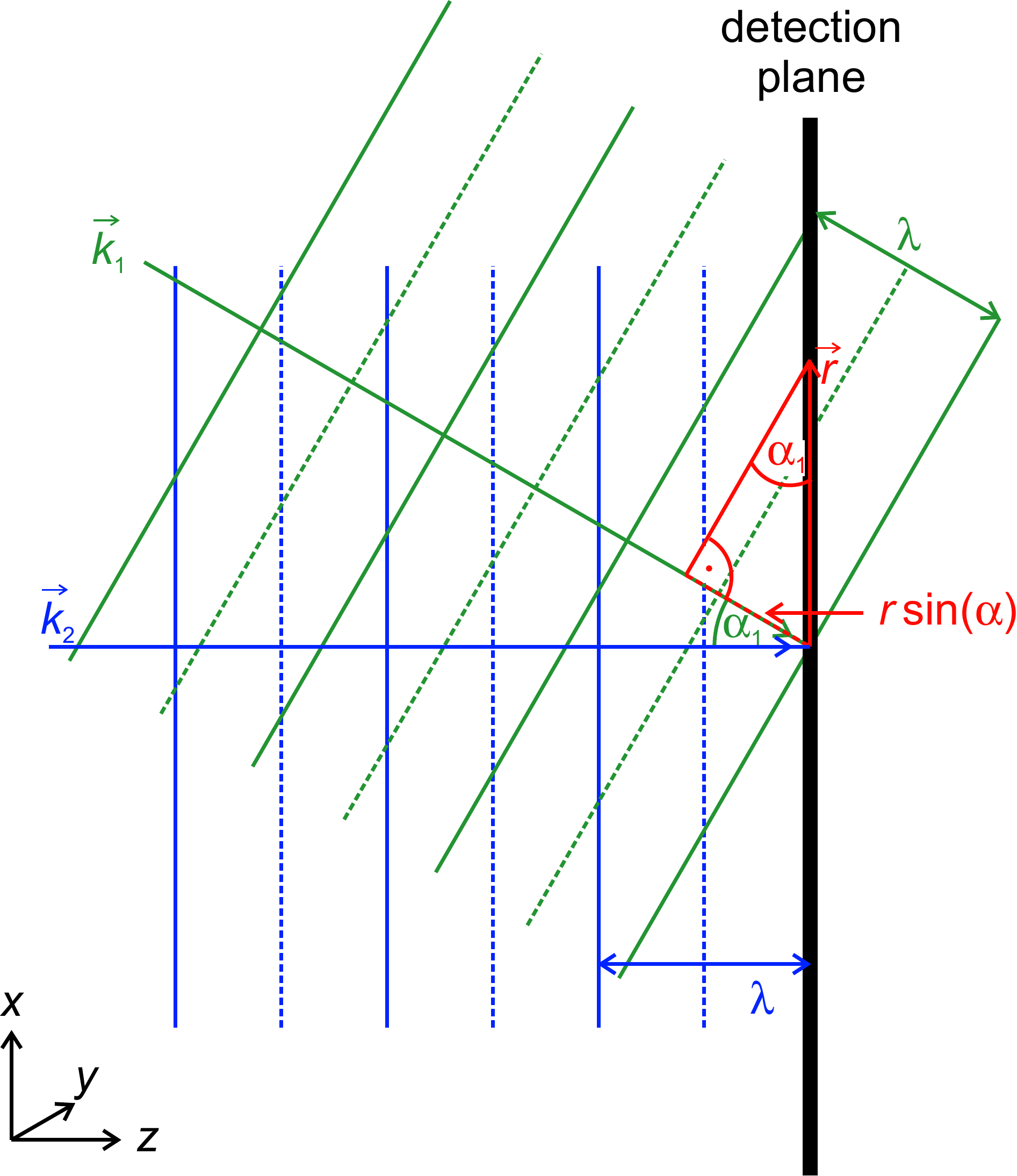}
	\caption{(color online) 
	Scheme of two superimposed, monochromatic plane waves with wavevectors $\vec{k}_1 = (2 \pi / \lambda) (- \sin (\alpha_1) \vec{e}_x + \cos (\alpha_1) \vec{e}_z) ,\vec{k}_2 = (2 \pi / \lambda) \vec{e}_z$ and equal wavelength $\lambda$ to illustrate the spatial phase distribution $\phi_{12}( \vec{r})$ (cf. second term in \eqr{eq:interf_phase}) for the special case of $\alpha_2 = 0$. The solid (dashed) lines represent electric field maxima (minima) of the corresponding waves. Interference maxima (minima) within the detection plane occur at position $\vec{r}$ for which the electric field maximum of wave 2 is superimposed with a field maximum (minimum) of wave 1. The spatial phase distribution in the detection plane $\phi_{12}(x,y)$ = $\phi_{12}(x) = (2 \pi |\vec{r}| \sin(\alpha_1)) / \lambda$ is in accordance with the second term in \eqr{eq:interf_phase} for $\alpha_2 = 0$. $\phi_{12}(x)$ can be found by relating the projection $|\vec{r}| \sin(\alpha_1)$ with the emission wavelength $\lambda$. This leads to a fringe period of $\Delta |\vec{r}| = \Delta x = \lambda /  \sin(\alpha_1)$. }
	\label{fig:scheme_PhaseDiff}
\end{figure}

The phase difference $\phi_{12}$ between the emission from both interferometer arms depends on the emission wavelength $\lambda$ in the following way:
\begin{align}
	\label{eq:interf_phase}
\phi_{12}(\lambda,\Delta s,\vec{r}) & =  \frac{2 \pi \Delta s} {\lambda} + (\vec{k_1} - \vec{k_2}) \vec{r} \notag \\ & = \frac{2 \pi \Delta s} {\lambda} +  \frac{2 \pi} {\lambda} |\vec{r}| (\sin(\alpha_1) - \sin(\alpha_2)).
\end{align}
Here, $\Delta s$ is the path length difference between both interferometer arms, $\vec{k_1},\vec{k_2}$ are the wavevectors of the waves propagating along the individual interferometer arms, $\alpha_1,\alpha_2$ are the angle between the optical axis and the corresponding wavevectors $\vec{k_1},\vec{k_2}$ and $\vec{r}$ is the distance vector from the intersection point between both wavevectors (in the detector plane) and the point of interest of the resulting interference pattern. The geometry is sketched in Fig.~\ref{fig:scheme_PhaseDiff} for the special case of $\alpha_2 = 0$. The second term of \eqr{eq:interf_phase} defines the appearance of the observed interference pattern (fringe distance and orientation) for a specified $\Delta s$ and $\lambda$, whereas the first term causes an additional, spectrally dependent phase offset that increases linearly with increasing path length difference.
The spectral resolution is $\Delta \lambda = \lambda_\mt{max} - \lambda_\mt{min} = 0.08$~nm for our experiment. Therefore, accumulation of the intensity of the interference pattern over the spectral range $\Delta \lambda$ is accompanied by an integration over a range of phase differences

\begin{align}
\label{eq:interf_phasediff}
\Delta \phi_{12}(\lambda,\Delta s) = & 2 \pi   \left(  \frac{1}{\lambda_\mt{min}} - \frac{1}{\lambda_\mt{max}} \right)  \notag \\ & (\Delta s + |\vec{r}| (\sin(\alpha_1) - \sin(\alpha_2)).
\end{align}

The second term in the second bracket in \eqr{eq:interf_phasediff} can be neglected due to $|\vec{r}| (\sin(\alpha_1) - \sin(\alpha_2)) \ll \Delta s$ for almost the total range of path differences used here. The angle between the propagation directions of the two superimposed waves is about $0.14^\circ$ deduced from the fringe distance of the observed interference pattern, which represents an upper limit for $\sin(\alpha_1) - \sin(\alpha_2) \approx \alpha_1 - \alpha_2$ (in case of opposite signs for both angles). The radius of the observed interference pattern is in the range of $|\vec{r}|_\mt{max} \approx 1$~mm leading to $|\vec{r}|_\mt{max} \Delta \alpha_\mt{max} \approx 2.4~\mu$m. This is about three orders of magnitude below the maximum path length difference of $\Delta s_\mt{max} = 2.7$~mm. 

The first term in \eqr{eq:interf_phasediff} can be re-expressed in terms of $\Delta \phi_{12}(\lambda,\Delta s) =  2 \pi   \left(  \frac{1}{\lambda_\mt{min}} - \frac{1}{\lambda_\mt{max}} \right) \Delta s = 2 \pi   \left(  \frac{\lambda_\mt{max} - \lambda_\mt{min}}{\lambda_\mt{max} \lambda_\mt{min}} \right) \Delta s \approx   2 \pi \frac{\Delta \lambda}{\lambda^2} \Delta s$. Therefore, a significant phase shift of the order of $2\pi$ is induced, if the path difference  $\Delta s$ is of the order of $\lambda^2 / \Delta \lambda$. This is the case for the measurement presented here, since $\lambda^2 / \Delta \lambda \approx 1.7$~mm. Thus, the experimentally observed decrease of the normalized visibility with increasing path difference (or temporal delay) is stronger than the pure reduction due to the decreasing temporal coherence $g^1(\Delta t)$, which is therefore under-estimated.

In order to quantify the impact of the spectrally dependent phase shift on the determined visibility we integrate the interference pattern $I_\mt{interf}$ over the range of phase differences $\Delta \phi_{12} = \phi_{12,\mt{max}} - \phi_{12,\mt{min}}$ that corresponds to a single CCD row
	\begin{align}
	I_\mt{interf}(\phi_{12,\mt{min}},\phi_{12,\mt{max}},\Delta s,\vec{r})  \notag \\ =  \int_{\phi_{12,\mt{min}}}^{\phi_{12,\mt{max}}} \! \frac{ I_\mt{interf}(\phi_{12},\Delta s,\vec{r}) \, \mathrm{d} \phi_{12}}{\phi_{12,\mt{max}} - \phi_{12,\mt{min}}}    
	\end{align} 
	where $\phi_{12}$ is given by Eq.~\ref{eq:interf_phase}. Thereby, we assume a constant intensity distribution of the emission from the individual arms within the width of the single CCD row as well as total coherence between both intensity signals. Afterwards, we determined the normalized visibility  according to Eq.~\ref{eq:Inorm} in the main text.
	\begin{figure}
		\centering
		\includegraphics[width=\linewidth]{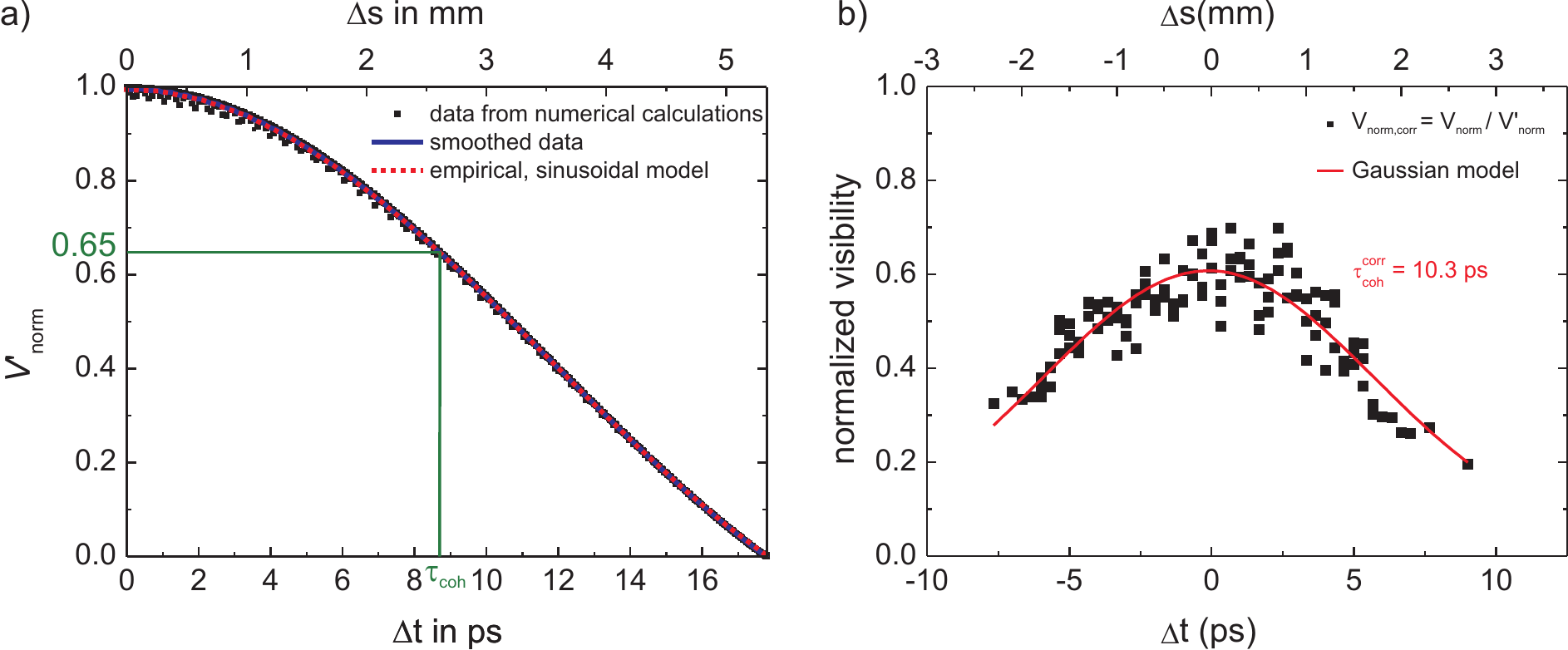}
		\caption{(color online)
			(a) Decrease of the normalized visibility with increasing temporal delay of a theoretically assumed, totally coherent signal due to the spectrally dependent phase shift. The small fluctuations of the numerically determined data are caused by the finite number of points for the spatial interference pattern $I_\mt{interf}(\vec{r})$. (b) Corrected normalized visibility $V_\mt{norm}'$ as a function of the temporal delay between both intensity signals.}
		\label{fig:SM_phaseShift_correction}
	\end{figure}
	Following this procedure, we found empirically a sinusoidal decrease of $V_\mt{norm}$ with increasing path length difference or rather temporal delay, as shown in Fig.~S\ref{fig:SM_phaseShift_correction}(a). By taking into account this systematic experimental error, a corrected coherence time of 10.3~ps could be deduced (cf. Fig.~S\ref{fig:SM_phaseShift_correction}(b)). 

\subsection{Limited Excitation Pulse Width}

For the coherence measurements we used pulsed excitation  with a pulse length of about 2~ps. Note that for the PL experiments which are compared to theoretical simulations based on a steady-state theory, a different excitation laser with pulse length of about 500~ps was used. By means of previous time-resolved measurements of the investigated microcavity (MC) under similar excitation conditions, a condensate lifetime of about 4-8 ps was observed (not shown in this work). However, the PL intensity decreases exponentially after the excitation pulse vanishes, which strongly limits the determination of longer coherence times.

\begin{figure}[htb]
\centering
\includegraphics[width=\linewidth]{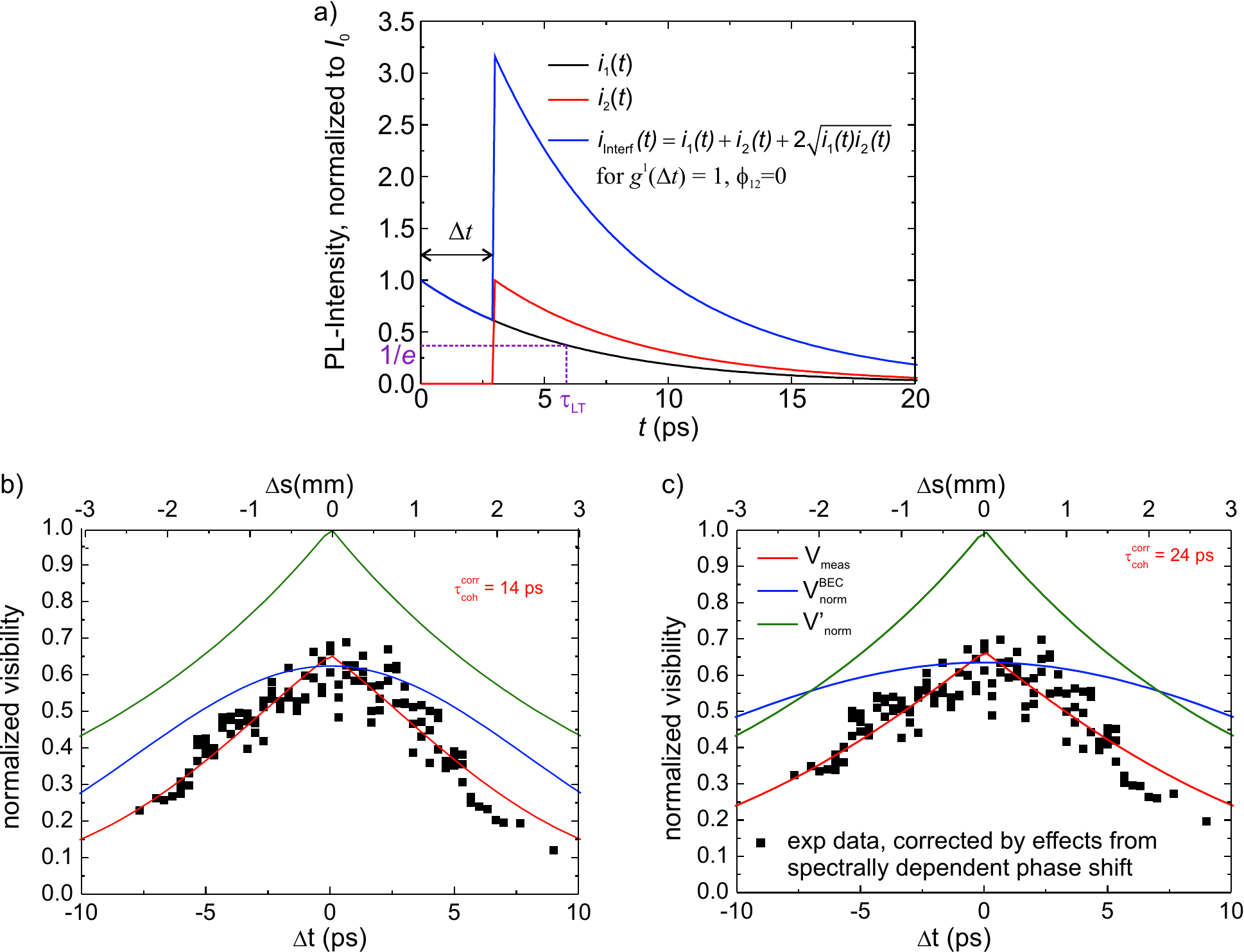}
\caption{(color online)
	(a) Monoexponential decay of two intensity signals $i_1(t)$ and $i_2(t)$ as well as of the superimposed interference signal $i_\mt{interf}(t)$. For the latter one, total coherence $g^1(\Delta t)$ and vanishing phase shift $\Delta \phi_{12}$ between both individual signals are assumed. (b) Comparison between the experimentally obtained normalized visibility and the adapted model $V_{\mt{meas}}  = V_\mt{norm}^\mt{BEC} \; V_{\mt{norm}}'$ (cf. Eq.~\ref{eq:tau_coh}.) Here, we considered the decrease of the calculated normalized visibility $V_{\mt{norm}}'$ or rather normalized temporal coherence with increasing temporal delay between both signals due to the finite pulse length of the condensate emission. We assumed a condensate lifetime of $\tau_\mt{LT} = 6~\mt{ps}$. (c) Similar procedure as in (b), but an additional correction due to the spectrally dependent phase shift was applied.}
\label{fig:SM_pulsedExc}
\end{figure}

To quantify the impact of the finite pulse duration on the calculated coherence time, we consider two pulses $i_1(t)$ and $i_2(t)$ originating from both interferometer arms with equal amplitude $i_0$ and with a temporal delay of $\Delta t$. For simplicity, we describe the temporal evolution of both pulses by a mono-exponential decay  

\begin{align}
\label{eq:exp_decay}
i_1(t) &= i_0 \:  \exp(-t/\tau_\mt{LT}) \\
i_2(t) &= \begin{cases} 0 & \text{for} \: t < \Delta t \\                         
			            i_0 \:  \exp(- (t+ \Delta t)/\tau_\mt{LT})  & \text{for} \: t \geq \Delta t  \end{cases}  
\end{align}  

while neglecting the onset time as shown in Fig.~\ref{fig:SM_pulsedExc}(a). 
For $t \geq \Delta t$ the intensity of the delayed signal $i_2(t)$ is by a factor of $A \equiv \exp(\Delta t / \tau_\mt{LT})$ larger than $i_1(t)$.  
The detection occurs time-integrated over millions of laser pulses, whereas each pulse acts as an individual statistical event. Thus, we have to consider the PL intensity integrated over the time interval between two consecutive pulses $T$:

\begin{align}
\label{eq:exp_decay_sum}
I_1 &= \int_0^T i_1(t) \: d\mt{t} & \overset{T \gg \tau_\mt{LT}}= i_0 \tau_\mt{LT}\\
I_2 &= \int_{\Delta t}^T i_2(t) \: d\mt{t}   & \overset{T \gg \tau_\mt{LT}}= i_0 \tau_\mt{LT} .
\end{align} 

The condition $T \gg \tau_\mt{LT}$ is fulfilled in our experiment, since the time interval $T$ is 13~ns, which is about three orders of magnitude larger than the lifetime of the condensate of about $\tau_\mt{LT} = 6~ \mt{ps}.$ To determine the coherence time, we calculate the normalized visibility $V_{\mt{norm}}$ of the interference pattern (cf.Eq.~\ref{eq:Inorm} in the main text), whose amplitude represents the temporal first order correlation function $g^1(\Delta t)$. This condition is only valid, if we calculate $V_{\mt{norm}}(t)$ for each point in time during the temporal decay of the PL intensity separately. But in our experiment, we firstly measure the temporally integrated intensity of the interference pattern $ I_{\mt{interf}} = \int_0^T  i_{\mt{interf}}(t) \: d\mt{t} = \int_0^T  (i_1(t) + i_2(t) + 2\sqrt{i_1(t) i_2(t)} g^1(\Delta t) \cos(\Delta \phi_{12}) ) \: d\mt{t}$  and calculate the normalized visibility $V_\mt{norm}'$ afterwards. To quantify the impact of the pulsed excitation we calculate $V_{\mt{norm}}'$ for an assumed superposition of two totally coherent signals without any phase shift ($g^1(\Delta t)~=~1,~\cos(\Delta \phi_{12} = 0$)).   This leads to the following equation:
\begin{align}
\label{eq:interf_tenp_integration}
I_{\mt{interf}}  &= \int_0^{\Delta t} i_1(t) \: d\mt{t} + \int_{\Delta t}^T  (i_1(t) + i_2(t) + \sqrt{i_1(t) i_2(t)} \: d\mt{t}  \notag \\
&= \int_0^{\Delta t} i_1(t) \: d\mt{t} + \int_{\Delta t}^T  
i_2(t) [\frac{1}{A} + 1 + \frac{2}{\sqrt{A}}] \: d\mt{t}  \notag  \\
&= i_0 \tau_\mt{LT} \: (1 - \frac{1}{A}) + i_0 \tau_\mt{LT} [\frac{1}{A} + 1 + \frac{2}{\sqrt{A}}] \notag \\
&= i_0 \tau_\mt{LT} \: (2 + \frac{2}{\sqrt{A}})  
\end{align} 
with $A$ being the intensity factor between both signals as defined previously in this paragraph.     
This value of $I_{\mt{interf}}$ is smaller than the expected maximum intensity for the superposition of two totally coherent signals with equal intensity of $I_{\mt{interf}}^{\mt{max}} = 4 i_0 \tau_\mt{LT}$, except for the trivial case of A = 1 that is fulfilled for $\Delta t  = 0$ only. Consequently, this leads also to a reduction of the normalized visibility 
\begin{align}
V_{\mt{norm}}' & = \frac{I_{\mt{interf}} - I_1 - I_2}
	{2\sqrt{I_1 I_2}}   \notag \\ & = \frac{1}{\sqrt{A}} = \exp(-\frac{\Delta t}{2 \tau_\mt{LT}})  & \leq 1 .
\end{align}  

Fig.~\ref{fig:SM_pulsedExc}(b) shows the evolution of the calculated normalized visibility $V_{\mt{norm}}'$  as a function of the temporal delay for $\tau_\mt{LT} = 6$~ps. Since we assumed total coherence and a vanishing phase shift between both signals $i_1(t)$ and $i_2(t)$, the reduction of $V_{\mt{norm}}'$ is exclusively caused by the temporal decay of the condensate emission due to the pulsed excitation. Therefore, $V_{{}\mt{norm}}'$ represents a correction function for the real value $V_{{}\mt{norm}}$. For the experimentally determined, uncorrected coherence time of $\tau_\mt{coh}' = 8.7~$ps we find a reduction of the normalized intensity by a factor of $V_{{}\mt{norm}}' = 0.49$. For comparison, the maximum value extracted from the experiment is about $V_{{}\mt{norm}} \approx 0.6$ (cf. Fig.~\ref{fig:EkSpace_Interference} in the main text), whereas for uncorrelated emission a residual normalized visibility of $V_{{}\mt{norm}} \approx 0.04$ could be estimated (not shown here). For our MC condensation can only be achieved with pulsed excitation, thus we cannot determine the real coherence time directly from the measurement. However, with the help of the simplified model presented here, a corrected value of the coherence time can be estimated. 

The normalized visibility obtained from the experiment $V_\mt{meas}$,  is a convolution of the normalized visibility of the investigated condensate $V_\mt{norm}^\mt{BEC}$, for which a Gaussian decay is assumed for increasing temporal delay $\Delta t$, and of the correction function $V_{\mt{norm}}'$, leading to the following equation:
\begin{align}
V_{\mt{meas}} & = V_\mt{norm}^\mt{BEC} \; V_{\mt{norm}}'  \notag \\ & 
= g^1(\Delta t =0) \exp(- \frac{\pi}{2} \frac{\Delta t^2}{\tau_{\mt{coh}}^2}) \exp(-\frac{\Delta t}{2 \tau_\mt{LT}})  \notag \\
 \label{eq:tau_coh} & =g^1(\Delta t =0) \exp(- \frac{\pi}{2} \frac{\Delta t^2}{\tau_{\mt{coh}}^2} - \frac{\Delta t}{2 \tau_\mt{LT}}).
\end{align}    

By comparing the experimental data with the corrected model obtained in Eq.~\ref{eq:tau_coh} we can estimate a corrected coherence time of $\tau_\mt{coh} = 14$~ps. If we further apply corrections from the spectrally dependent phase shift, as discussed in the previous section, a maximum coherence time of $\tau_\mt{coh} = 24$~ps was estimated.

\end{document}

%% file: source_v2.bbl
%